\documentclass[pre,aps,twocolumn,showpacs]{revtex4}
\def\bxi{\mbox{\boldmath$\xi$}}

\def\btau{\mbox{\boldmath$\tau$}}
\usepackage{graphicx}
\begin{document}
\title
{Theory of self-assembly  of microtubules and motors}
\author{Igor S. Aranson$^1$ and Lev S. Tsimring$^2$}
\affiliation{$^1$Argonne National Laboratory, 9700 South Cass
Avenue, Argonne, Illinois, 60439\\
$^2$Institute for Nonlinear Science, University of California, San
Diego, La Jolla, CA 92093-0402}
\date{\today}

\begin{abstract}
We derive a model describing spatio-temporal organization of an
array of microtubules interacting via molecular motors. Starting
from a stochastic model of inelastic polar rods with a generic
anisotropic interaction kernel we obtain a set of  equations for
the local rods concentration and orientation. At large enough mean
density of rods and concentration of motors, the model describes
an orientational instability. We demonstrate that the orientational
instability leads to the formation of vortices and (for large
density and/or kernel anisotropy) asters seen in recent
experiments. We derive the specific form of the interaction kernel from
the detailed analysis of microscopic interaction of two filaments
mediated by a moving molecular motor, and extend our results to
include variable motor density and motor attachment to the substrate.
\end{abstract}

\maketitle

\section{Introduction}
\label{intro}
One of the most important functions of molecular
motors is to organize a network of long filaments (microtubules)
during cell division to form cytosceletons  of daughter cells
\cite{howard00}. In order to better understand the details of this
complex self-organization process, a number of  {\em in vitro}
experiments were performed
\cite{takiguchi91,urrutia91,nedelec97,surrey01,humpgrey02,nedelec01}
to study interaction of molecular motors and microtubules   in
isolation from other biophysical processes simultaneously
occurring {\em in vivo}. At large enough concentration of
molecular motors and microtubules, the latter organize in {\em
asters} and {\em vortices} depending on the type and concentration
of molecular motors.

\begin{figure}[ptb]
\includegraphics[width=3.3in,angle=0]{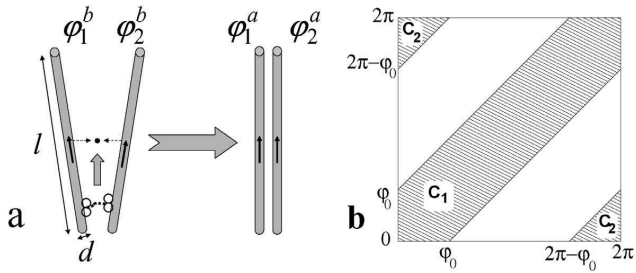}
\caption{a - sketch of motor-mediated two-rod interaction for
$\gamma=1/2$, b - integration regions $C_{1,2}$ for
Eq.(\protect\ref{master}).} \label{fig:sketch}
\end{figure}

In the above experiments the elementary interaction processes were
identified.  After molecular motor binds to a microtubule at a
random position, it marches along it in a fixed direction until it
unbinds without appreciable displacement of microtubules (since
the mass of a molecular motor is small in comparison with that of
the microtubule).  However, if a molecular motor binds to {\em
two} microtubules (most molecular motors have at least two binding
sites), it can change their mutual position and
orientation significantly. In small-scale simulations
\cite{surrey01}, the interaction of rod-like filaments via motor
binding has been studied, and patterns resembling
experimental ones were observed. In \cite{lee01} a
phenomenological model for the molecular motor density  and the
microtubule orientation has been proposed. The model included
transport of molecular motors along microtubules and alignment of
microtubules mediated by molecular motors. Simulations showed that
vortices and asters indeed form in this model, however,  only one large
vortex formed  in case of high density of motors. Ref.
\cite{kim03} generalized this model by including separate
densities of free and bound molecular motors, as well as the
density of microtubules. This model exhibited a transition from asters to
vortices as the density of molecular motors is increased,  in
apparent disagreement with experimental evidence \cite{nedelec01} that
asters give way to vortices with {\em decreasing} the molecular
motors concentration. Somewhat similar approach was employed more
recently by Sankararaman et al. \cite{sank}, again with the same
apparent disagreement with the experiment as in Refs.
\cite{lee01,kim03}. A phenomenological flux-force relation for
active gels was introduced in \cite{kruse}. While vortex and aster
solutions were obtained in a certain limit, an analysis of that
model is difficult because of the large number of unknown parameters
and fields.

In Ref. \cite{marchetti03} a set of equations for microtubules
density and orientation was  derived from conservation laws for
microtubules probability distribution function. These conservation
laws were based on the phenomenological expressions for the
probability fluxes due to diffusion and motor-mediated
interactions. The latter, however, assumed that tubules are only
displaced and rotated infinitesimally in individual interactions,
which may not be the case in experiments. The model does not
produce onset of spontaneous orientation for any density of
microtubules. The authors argue that asters and vortices may be
created as a result of the ``bundling instability'' \cite{zimm}.
However, this model demonstrated the bundling instability at small
densities of tubules, and oscillatory orientational instability at
large densities contrary to observations in which the
orientational ordering is observed at smaller densities and is not
oscillatory. In subsequent publication \cite{marchetti05}, a
derivation of the probability conservation equations from
microscopic mean-field model of forces between tubules and motors
was presented, however the assumptions made in the course of
derivation (i.e. infinitely stiff molecular motors)  lead to a
surprising conclusion that filaments do not change their
orientation during interaction.

In this paper we present an alternative calculation of the
microscopic motion of two filaments connected by a  moving motor
in a viscous fluid and show that filaments do change orientation
as a result of the interaction. In our short publication
\cite{at_short}  we derived  a continuum model for the collective
spatio-temporal dynamics of microtubules starting with a
stochastic microscopic master equation for interacting inelastic
polar rods, assuming that the density of molecular motors is
homogeneous in space.  Our model differed from the transport
equations \cite{marchetti03} in that it treated the interaction
between two tubules as an instantaneous inelastic ``collision''
that can change the orientation of the filaments significantly.
The model exhibits an onset of orientational
order for large enough density of mictorubules and molecular
motors \cite{ahmadi05}, formation of vortices and then asters with increase in the
molecular motors concentration, in a qualitative agreement with
experiment.

In this paper we present a more detailed derivation of our equations
starting from a microscopic model of tubule-motor interaction. We
derive the interaction kernel from microscopic rules
and relate the kernel characteristics with the properties of the
motors. We further extend our analysis: we lift the assumption of
the homogeneous distribution of molecular motors and include an
additional equation for the evolution of  the molecular motor
density. We also consider the situation when some molecular
motors are attached to a substrate (usually, a glass plate). These
modifications allow as to improve an agreement with experiment.
We explain accumulation of molecular motors at the center of
an aster by advection along microtubules and slow rotation of
vortices by the interaction of microtubules with the attached
motors, as it was observed in experiments.

The structure of the paper is as follows. In Sec. \ref{sec1} we
present derivation of the coarse-grained equation for orientation
and density from the microscopic Maxwell model for interacting
polar rods. In Sec. \ref{int_kernel} we calculate the form of the
interaction kernel from microscopic rules of interaction of two
filaments. In Sec. \ref{sec2} the effects of spatial coupling are
considered. First in Sec. \ref{cont} we reduce stochastic
equations for the probability function to the set of equations for
local orientation and density of micortubules. In Sec. \ref{sec3}
the stability analysis of the isolated vortex and aster solutions
is performed and the phase  diagram of various regimes is
presented as the function of motor density and anisotropy of the
interaction kernel. In Sec. \ref{sec4} we include effects of
motors attached to the substrate and explain rotation of the
vortices and onset of large variations of the microtubules
density. In Sec. \ref{sec5} we consider effects of variable motor
density and derive the equation for the evolution of the motors.
Technical details of derivations are presented in Appendices.

\section{Maxwell model for inelastic polar rods}
\label{sec1}

At this stage, molecular motors enter the model implicitly by specifying the
microscopic interaction rules between two rods. Since the
diffusion of small motors is about two order of magnitude higher
than that of large and heavy  mictortubules, in this Section we neglect
spatial variations of the motor density and treat the collision rules
as spatially homogeneous. Effects of variable motor
density are considered in Sec. \ref{sec5} below.  Each rod is assumed
to be of length $l$ and diameter $d\ll l$, and is characterized
by the position of its center of mass ${\bf r}$ and orientation
angle $\phi$.

Consider the orientational dynamics only and ignore the locales
of interacting rods (an analog of the Maxwell model of
binary collisions in kinetic theory of gases, see e.g.
\cite{bennaim00}). We model the motor-mediated inelastic
interaction by an instantaneous collision in which two rods change
their orientations according to the following collision rule:
\begin{equation}
\left(\begin{array}{c} \phi^a_1 \\ \phi^a_2
\end{array}\right) = \left( \begin{array}{cc} \gamma   & 1-\gamma  \\
1-\gamma &  \gamma \end{array} \right) \left(\begin{array}{c} \phi^b_1
\\ \phi^b_2
\end{array}\right)
\label{collis}
\end{equation}
where $\phi_{1,2}^b$ are the two rods' orientations  before and $\phi_{1,2}^a$
after the collision, and  $\gamma$ characterizes inelasticity of
collisions. The angle
between two rods is reduced after the collision by a factor
$2\gamma-1$. Totally elastic collision corresponds to $\gamma=0$
(the rods exchange their angles) and a totally inelastic collision
corresponds to $\gamma=1/2$: rods acquire identical orientation
$\phi^a_{1,2}=(\phi^b_1+\phi^b_2)/2$ (see Fig.
\ref{fig:sketch},a). Here we assume that two rods only interact if
the angle between them before collision is less than $\phi_0$,
$|\phi_2^b-\phi_1^b|<\phi_0<\pi$.
Because of $2\pi$-periodicity, we have to add the rule of collision between
two rods with $2\pi-\phi_0<|\phi^b_2-\phi^b_1|<2\pi$.  In this
case we have to replace $\phi^{b,a}_1\to\phi^{b,a}_1+\pi,
\phi^{b,a}_2\to\phi^{b,a}_2-\pi$ in Eq. (\ref{collis}). In the
following we will only consider the case of totally inelastic rods
($\gamma=1/2$) and $\phi_0=\pi$, the generalization for arbitrary
$\gamma$ and $\phi_0$ is straightforward, see Appendix \ref{app1}.
The probability $P(\phi)$ obeys the following master equation

\begin{figure}[ptb]
\includegraphics[width=3in,angle=0]{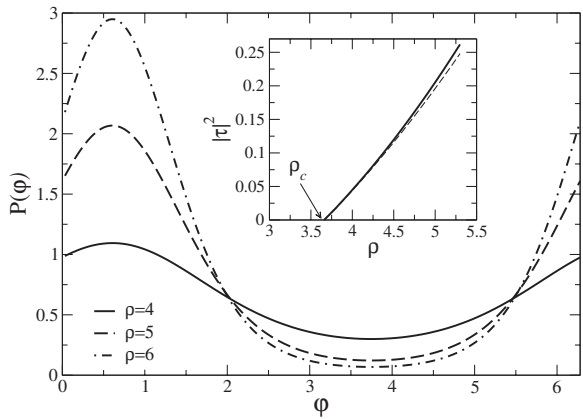}
\caption{ Stationary solutions $P(\phi)$ for different $\rho$.
Inset: the stationary value of $|\btau|$ vs  $\rho$  obtained from
the Maxwell model (\protect\ref{master2c}), dashed line -
truncated model (\protect\ref{Pk2c}).} \label{fig:distr}
\end{figure}
\begin{eqnarray}
&&\partial_t P(\phi)=
D_r\partial^2_\phi P(\phi)
 +g \int_{C_1} d\phi_1d\phi_2P(\phi_1)P(\phi_2) \label{master}\\
 &&\times[\delta(\phi-
\phi_1 /2 -\phi_2/2  )
 -\delta(\phi-\phi_2)]+g \int_{C_2} d\phi_1d\phi_2
 \nonumber \\
 &&\times P(\phi_1)P(\phi_2)
 [\delta(\phi-\phi_1 / 2 - \phi_2/2 -\pi)-\delta(\phi-\phi_2)]
\nonumber
\end{eqnarray}
where  $g$ is the ``collision rate'' proportional to the number
density of molecular motors $m$, the diffusion term $\propto D_r$
describes thermal fluctuations of rod orientations, and the
integration domains $C_1, C_2$ are shown in Fig.\ref{fig:sketch}b.
From the dimensional analysis one finds that the collision rate
$g$ is of the order of $m D_r S_0$, since $1/D_r$ is the only time
scale in Eq. (\ref{master}), and $S_0 \sim l^2$ is the interaction
cross-section of microtubules.

Changing variables $t \to D_r t$, $P \to  g P/D_r $, $
w=\phi_2-\phi_1$, one obtains
\begin{eqnarray}
&&\partial_t P(\phi)
=\partial^2_\phi P(\phi) + \int_{-\pi}^{\pi}dw
\nonumber \\
&&\times \left[P(\phi+w/2)P(\phi-w/2) -P(\phi)P(\phi-w)\right]
\label{master2c}
\end{eqnarray}
The rescaled number density
 $\rho=
\int_0^{2\pi}P(\phi,t)d\phi$ now is proportional to the {\it density
of rods multiplied by the density of motors}. In the following,
an increase of the density of molecular motors is reflected in our
analysis as an increase of the number density  $\rho$.

\subsection{Orientation instability}

Eq. (\ref{master2c}) possesses uniform steady-state solution
$P(\phi)=P_0=\rho/2 \pi=const$ corresponding to isotropic
distribution of rods. This solution looses its stability with
respect to anisotropic perturbations with the increase of density
$\rho$. The instability signals the onset of spontaneous
orientation. Substituting  solution to Eq. (\ref{master2c}) in the
form $P(\phi,t) = P_0+ \xi(\phi,t)$, where $\xi$ is small
perturbation, we obtain linear equation for $\xi$:
\begin{eqnarray}
\partial_t \xi(\phi)&=& \partial^2_\phi \xi(\phi)+
\frac{\rho}{2 \pi}
 \int_{-\pi}^{\pi} \left(\xi(\phi+w/2) \right. \nonumber
 \\
  &+& \left.\xi(\phi-w/2)-\xi(\phi-w)-\xi(\phi)  \right) d w
\label{master_lin}
\end{eqnarray}
Looking for the solution to Eq. (\ref{master_lin}) in the form
$\xi \sim \exp [ \lambda_k t \pm  i k \phi]$, where $k\ne 0$ is integer,
for the growth rate $ \lambda_k$ we find
\begin{equation}
\lambda_k =\rho \left( \frac{4}{k \pi} \sin (\pi k/2)-1\right)  -
k^2 \label{linear1}
\end{equation}
Thus, it follows from Eq. (\ref{linear1}) that perturbations
with $k = \pm 1$ have the largest growth rate $\lambda_1 = \rho
(4/\pi -1) -1$. The instability ( $\lambda_1 >0$ ) occurs for the
density $\rho> \rho_c = \pi /(4-\pi) \approx 3.662 $, and leads to
breaking the azimuthal symmetry and formation of anisotropic, i.e.
oriented states. The resulting orientation is determined by initial
conditions contained in the perturbation $\xi$.

\subsection{Fourier expansion}
Let us consider the Fourier harmonics of the probability density
$P(\phi)$:
\begin{equation}
P_k=\langle e^{-ik\phi}\rangle=\frac{1}{2 \pi} \int _0^{2 \pi} d
\phi e^{-i k \phi} P(\phi,t) \label{cumm}
\end{equation}
The zeroth harmonic $P_0=\rho/2\pi=const$, and the real and
imaginary parts of $P_1$ represent the components $\tau_x= \langle
\cos \phi\rangle, \tau_y=\langle \sin \phi\rangle$ of the average
orientation vector $\btau$, $\tau_x+i\tau_y=P_1^*$.  Substituting
 (\ref{cumm}) into Eq.(\ref{master2c}) yields:
\begin{equation}
\dot P_k +( k^2+\rho)P_k=2\pi \sum_{m} P_{k-m} P_m S[\pi k/2- m
\pi] \label{Pk1a}
\end{equation}
(here $S(x)=\sin x/x$). Due to the rotational diffusion term, the
magnitudes of high-order harmonics decay exponentially with $|k|$,
see Eq. (\ref{linear1}). Neglecting all $P_k$  for  $|k|>2$ we
obtain from Eq.(\ref{Pk1a})
\begin{eqnarray}
\dot P_1 &+& P_1= P_0 P_1 2 (4 -\pi) -{8\over 3}  P_2 P_1^*
\label{Pk2} \\
 \dot P_2 &+&4P_2= -P_0 P_2 2\pi  + 2 \pi
P_1^2 \label{Pk3}
\end{eqnarray}

Since near the instability threshold the decay rate of $P_2$ is
much larger than the growth rate of $P_1$, see Eq.
(\ref{linear1}), we can neglect the time derivative $\dot P_2$ and
obtain $P_2=AP_1^2$ with $A=2\pi(\rho+4 )^{-1}$ and arrive at:
\begin{equation} \dot{\btau} =\epsilon \btau -A_0|\btau|^2\btau
\label{Pk2c}
\end{equation}
with
\begin{eqnarray}
\epsilon&=&\rho  (4\pi^{-1}-1)-1\approx 0.273 \rho -1 \nonumber
\\
A_0&=& 8 A/3=\frac{16 \pi}{3 (\rho+4 )} \label{eps1}
\end{eqnarray}
For large enough $\rho
>\rho_{c}=3.662$, an ordering instability
leads to spontaneous rods alignment. This
instability saturates at the value determined by $\rho$. Close to
the threshold $A_0 \approx 2.18$.

In order to verify our approximations we solved Eq.
(\ref{master2c}) for $\gamma=1/2, \phi_0=\pi$ numerically
by the finite difference method.
We find that random initial conditions rapidly evolved
towards a single-peaked stationary distribution, the position of
the maximum of the distribution being determined by initial
conditions. Fig. \ref{fig:distr} shows typical  stationary
solutions $P(\phi)$ obtained from Eq. (\ref{master2c}) for
different values of $\rho$. One sees that $P(\phi) $ is weakly
non-uniform near the critical density, and becomes more peaked
with the increase of the
density.  From the numerically obtained distribution we extracted
orientation amplitude $|\btau|$ and compared it with the
analytical result $|\btau|^2=\epsilon/ A_0$ from Eqs.
(\ref{Pk2c}),(\ref{eps1}).
 As seen in the Inset, the corresponding values of
$|\btau|$ are consistent with the truncated model (\ref{Pk2c}) up
to $ \rho <5.5$.

We also studied Eq. (\ref{master2c}) for $\phi_0<\pi$. Whereas for
$\phi_0$ close to $\pi$ no qualitative difference was found, for
smaller $\phi_0$, e.g. $\phi_0<\pi/2$ we often obtained
long-living multiple peak distributions, with the number of peaks
roughly $\pi/\phi_0$. While the distribution possibly relaxes
towards a single peak, the transient time appears to be very large
due to exponentially weak interaction between the peaks.

\section{Microscopic picture of tubule-tubule interaction}
\label{int_kernel}
In order to generalize our model to account for spatial localization of
interaction between tubules, we need to introduce a more specific model
of interaction between two tubules mediated by a molecular motor.
Namely,
we should specify the ``collision rules'', or a relationship between
positions and orientations of two tubules before and after the interaction
via motor attachment/detachment, and the ``collision rate'', or the
probability of the collision to occur given the positions and
orientations of two tubules. The latter will play a role of interaction
kernel in the corresponding master equation for the tubule probability
distribution.

\subsection{Collision rules}
\begin{figure}[ptb]
\includegraphics[width=3. in,angle=0]{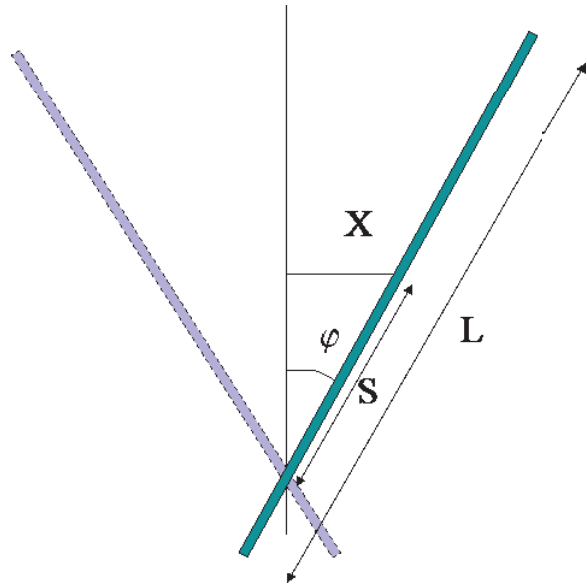}
\caption{Sketch  of the interaction of two microtubules }
\label{Fig1_kern}
\end{figure}

Here we specify these rules by integrating the equations of motion
of the two tubules.  This calculation is based on a number of
simplifications. We assume that two infinitely rigid rods of equal
length $l$ interact with one molecular motor. We assume that the
motor moves with constant speed $V$ along the rods (the results
can are trivially generalized for the case of $V \ne const$). To
simplify the system even further, we consider a symmetric case:
the distance of the motor from the center point of the rod $S$,
$-l/2<S<l/2$ is the same for both rods, see Figure
\ref{Fig1_kern}. Since the size of a motor ($\approx 30$ nm) is
much smaller than the length of a microtubule ($5..10$ micron), we
consider a limit of zero motor size. Since the motor's bending
elasticity is rather small,   we approximate the motor by a soft
spring and prescribe that the force $F$ exerted on the tubules due
to motor motion is perpendicular to the bisector of the angle
between the tubules (i.e. along the motor), which in the symmetric
case is along the $x$-axis, see Fig. \ref{Fig1_kern}. Even if the
symmetry is initially broken, and the force is exerted at an angle
to the $x$-axis, the force will initiate a relative displacement
of the tubules in the $y$ direction which will shift the binding
points in such a way as to restore the symmetry.

The equations governing evolution of the angle $\phi$ between the
microtubule and the bisector and the coordinates $X,Y$ of the
center of mass of the microtubule are obtained from balance of
torques and forces due to motor motion and viscous drag forces
\begin{eqnarray}
\partial_t \phi&=&  \xi_r^{-1} S \cos(\phi ) F \\
\partial_t X & =&  (\xi_\parallel^{-1}\cos^2\phi+\xi_\perp^{-1}\sin^2\phi) F, \\
\partial_t Y&=&(\xi_\parallel^{-1}-\xi_\perp^{-1})\sin\phi\cos\phi F
\label{rods1a}
\end{eqnarray}
Here $\xi_r,\xi_\parallel, \xi_\perp$ are rotational and
translational viscous drag coefficients, see Eq. (\ref{drug})
below. In the following we neglect the anisotropy of the
translation friction ($\xi_\parallel=\xi_\perp$ \cite{friction}),
then the equations will simplify considerably
\begin{eqnarray}
\partial_t \phi&=&  \xi_r^{-1} S \cos(\phi ) F \\
\partial_t X & =&  \xi_\parallel^{-1} F, \\
\partial_t Y&=&0.
\label{rods1}
\end{eqnarray}

Additional kinematic equation is obtained  from the condition that
the motor is attached at the distance $S$ from the center of tubule,
which gives
\begin{eqnarray}
X= -S  \sin  \phi \label{rods2}
\end{eqnarray}

Differentiating  Eq. (\ref{rods2}) with time and using $dS/dt=V$,
we exclude $F$ and derive an equation for $\phi$ (note that the
analysis in this Section is also valid for arbitrary
time-dependent velocity of the motor $V(t) >0$)
\begin{eqnarray}
{d\phi\over dt}&=& - \frac{ k V  S\cos( \phi )\sin( \phi)
}{1 + k S^2 \cos^2 (\phi ) } \label{phi2}
\end{eqnarray}
where $k=\xi_\parallel/\xi_r \approx 12 /l^2$, see Eq. (\ref{drug}).  We make
the following substitutions:
\begin{equation}
\tau \to kS^2    ; u \to \cos (\phi)^2 \end{equation}

In new variables Eq (\ref{phi2}) can be written as
\begin{equation}
\frac{ d \tau} {d u} = \frac{1+ \tau u}{u(1-u)} \label{rods4}
\end{equation}
Eq. (\ref{rods4})  is linear with respect to $\tau$ and
therefore has an exact solution
\begin{equation}
\tau= \frac{C+\log(u)}{1-u}
\end{equation}
where $C$ is a constant determined by initial conditions.
Returning to original variables, we obtain
\begin{equation}
\frac{C+\log(\cos^2 (\phi))}{\sin(\phi)^2}= k S^2 \label{phi3}
\end{equation}

For small angles $\phi$  Eq. (\ref{phi3}) simplifies and we obtain
\begin{equation}
\phi=\phi_0 \frac{\sqrt{1+k S_0^2 }}{\sqrt{1+k S^2}}
\end{equation}
where $\phi_0$ and $S_0$ are the initial conditions at $t=0$.

For the final angle obtained when the motor reaches  the end of
the microtubules ($S=l/2$) we obtain
\begin{equation}
\tilde \phi=\phi_0 \frac{\sqrt{1+k S_0^2 }}{\sqrt{1+kl^2/4}}
\label{rods5}
\end{equation}

As one sees from Eq. (\ref{rods5}), the final angle $\tilde \phi$
depends on the initial angle $\phi_0$ and the initial attachment
position $S_0$. Assuming that the probability of attachment of
the motor is independent of the position along the microtubule
$S$, in the small angle approximation for average angle
\begin{equation}
\langle \tilde \phi \rangle=l^{-1} \int_{-l/2}^{l/2} \tilde
\phi(S_0)  d S_0 \label{aver}
\end{equation}
we obtain
\begin{equation}
\langle \tilde \phi \rangle=\phi_0 \left[
\frac{1}{2}+\frac{\mbox{asinh}(\sqrt{k
l^2}/2)}{\sqrt{kl^2}\sqrt{1+kl^2/4}} \right]
\end{equation}
Thus, the averaged change in the angle  $\epsilon=\langle \phi
\rangle / \phi_0 $ is
\begin{equation}
\epsilon=\frac{1}{2}+\frac{\mbox{asinh}(\sqrt{kl^2}/2)}{\sqrt{kl^2}\sqrt{1+kl^2/4}}
\end{equation}
Obviously, for $k l^2 \to \infty$ the relative angle change $\epsilon
\to 1/2 $. Correspondingly, the restitution coefficient $
\gamma=(1 -\epsilon)/2  \to 1/4$, in agreement with our
assumptions on inelastic collision between the rods. For the case
of rigid rods we obtain from Eq. (\ref{drug}) that $k l^2
\approx 12$, which gives $\epsilon\approx 0.67$, or $\gamma\approx
0.17$, which is considerably smaller than fully-inelastic case of
$\gamma=1/2$. With the increase of $k$ the coefficient $\gamma$
increase, e.g. for $k=30$ one obtains $\gamma=0.2$.

\begin{figure}[ptb]
\includegraphics[width=2.7in,angle=-90]{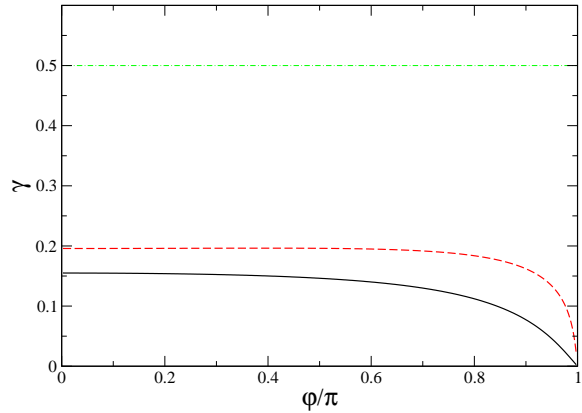}
\caption{The effective restitution coefficient $\gamma$ calculated
from Eq. (\ref{aver})  for $k=12$ (solid line). For comparison
$\gamma$ at $k=30$ is also shown (dashed line). Dot-dashed line
indicates the limit of fully inelastic interaction $\gamma=0.5$.  }
\label{Fig3_kern}
\end{figure}

For arbitrary $\phi_0$, in order to find the average angle change we solved
Eq. (\ref{phi3}),(\ref{aver})  numerically (see Fig.
\ref{Fig3_kern}). There is a very week dependence of
$\gamma$ on the initial angle $\phi$.
In a wide range of angles $\Delta \varphi <0.75 \pi$, parameter
$\gamma \approx 0.2 $ and then
$\gamma \to 0$ for $ \phi \to \pi$.

\subsection{Collision rate}
\label{sec:kern}
Now we turn to the calculation of the collision rate
$W({\bf r}_1,{\bf r}_2,\phi_1,\phi_2)$ between two tubules
with center-of-mass positions ${\bf r}_1, {\bf r}_2$ and orientation angles $\phi_1,\phi_2$.
We consider a translationally and rotationally invariant
system, so the collision rate depends only on the position and
orientation differences, $W({\bf r},\phi)$, where ${\bf r}={\bf r}_1-
{\bf r}_2$, and, as before, $\phi=(\phi_1-\phi_2)/2$.

Microtubules interact via a motor attachment only if they
intersect, which is expressed by the condition
\begin{equation}
|{\bf r_1} - {\bf r_2}| < \frac{l}{ 2}  \sin\phi
\end{equation}
This gives the collision rate in the form
\begin{equation}
W= g\Theta \left( |{\bf r_1} - {\bf r_2}| - l/2  \sin\phi\right)
\label{W1}
\end{equation}
where $\Theta$ is the $\Theta$-function, and $g$ is the
probability of a motor to bind to both tubules given that they
intersect. The latter is directly proportional to the
concentration of molecular motors bound to a tubule.  If the
concentration of the motors along the tubules were uniform, the
collision rate would be uniform inside the parallelogram defined
by Eq.(\ref{W1}).  However, due to the transport of motors along
the tubules, their concentrations increase towards the polar ends
of the filaments.

In order to compute the inhomogeneous motor distribution along a
filament, we
assume that the motors can be in two different states, bound and free.
The concentration of free motors $m_f$ is a function of
the coordinate along the tubule $S$, ($-l/2<S<l/2$),
perpendicular coordinate  $r_\perp$ (we  consider
a two-dimensional domain), and time $t$. Bound motors are localized on the tubule
itself, so their concentration is $M_b(S,t)\delta(r_\perp)$.

The inhomogeneous motor distribution
can be evaluated  from the following microscopic equations for
motor attachment/detachment and advection processes:
\begin{eqnarray}
\partial_t m_f & =&  (p_{off} M_b-p_{on} m_f)\delta(r_\perp)  + D \nabla^2 m_f  \\
\partial_t M_b & =&  -p_{off} M_b +p_{on} m_f(S,0,t) -\partial_S(V
M_b) \nonumber \label{motor-1}
\end{eqnarray}
These equations, formulated in terms of the concentration of
bound/free motors $M_b,m_f$, describe random binding/unbinding of
the motors with the probabilities   $p_{on,off}$; diffusion of free
motors (diffusion coefficient $D$), and the advection of
bound motors with the velocity $V$ along the tubule.  The
parameter $p_{off}$ characterizes processivity of molecular motors:
large $p_{off}$ corresponds to small processivity, motor unbinds
soon after it binds to a filament.

According to experiments (see e.g. \cite{case1997}),  while
multiple motors can be attached to a single tubule, only one motor
can be attached per elementary binding site (which represents a
section of approximately $l_0=10$ nm along the tubule). This leads
to a kind of hard-core repulsion which in the simplest
approximation can be taken into account by introducing local
``pressure'' $P$ of bound motors, and modifying the transport
speed $V$,
\begin{equation}
V=V_0-\eta\partial_s P
\end{equation}
where $\eta$ is an effective mobility. Pressure $P$ diverges as as the
bound motor density approaches densely packed limit of one motor per binding site
$M_b\to M_0 \sim 1/l_0$. We will
adopt a simple generic expression for the pressure as a function of the bound motor density
(see for comparison expressions for pressure in granular
hydrodynamics near closed-packed density
\cite{bennaim,haff,jenkins})
\begin{equation}
P=\frac{M_b T}{1-M_b/M_0}  \label{pressure}
\end{equation}
The ``temperature'' $T$ is determined by fluctuations  of bound
motors on the tubule and is typically  small, so the pressure can be
neglected everywhere except where the density is very close to the dense
limit.

In the stationary state Eqs. (\ref{motor-1}) assume the form
\begin{eqnarray}
(p_{off} M_b&-&p_{on} m_f)\delta(r_\perp)  + D \nabla^2 m_f=0
\label{motor-2a} \\
 -p_{off} M_b &+&p_{on} m_f(S,0)  =\partial_S \left[V_0  M_b-
\eta M_b\partial_SP \right] \label{motor-2b}
\end{eqnarray}

Since the diffusion constant $D$ of free motors is large, we can
neglect the inhomogeneity in the free motor distribution and
assume $m_f=const$ in Eq.(\ref{motor-2b}). Eq. (\ref{motor-2b})
has to be solved with the boundary condition $M_b=0$ at $S=-l/2$.
At the end of the tubule $S=l/2$ the ``exit'' flux of bound motors
$VM_b$ is determined by the detachment probability $p_{end}$ of
the motor, resulting in the condition $V M_b=p_{end} l_0 M_b$. For
small $T$, the density of bound motors has two distinct phases:
low-density ``gas phase'' near the beginning of the tubule
($S=-l/2$), and a high density ``solid phase'' near the end
($S=l/2$) \cite{cell}.
 The location of the boundary between these phases
can be found by equating the fluxes of bound motors in the two
phases. In the low density phase, the pressure term can be
neglected due to small $T$, and the solution has the form
\begin{equation}
M_b={p_{on}m_f\over p_{off}}\left[1-\exp\left(-{p_{off}\over V_0}
(S+l/2)\right)\right]
\end{equation}
Typically, $p_{off}V_0^{-1} l\gg 1$, so the density saturates very
quickly to the equilibrium value $M_e=p_{on}p_{off}^{-1}m_f$. This
solution corresponds to a constant flux of motors along the
tubule,
\begin{equation}
F_1=V_0 M_e\label{F1}=\frac{V_0 p_{on} m_f } {p_{off}}
\end{equation}

In the solid phase at a very low temperature $T$, the motor
density is very close to $M_0$. Thus, at the end of
the tubule, the flux of motors is equal to the number of motors
leaving the tubule in a unit time $p_{end}M_0$. According to
Eq.(\ref{motor-2b}), the flux of motors
in the solid phase is a linear function of the coordinate,
\begin{equation}
F_2=p_{end}l_0 M_0+\left[ p_{off} M_0-p_{on} m_f \right](l/2-S)
\end{equation}
The two phases are separated by a narrow interface
(the width of the interface vanishes when $T \to 0 $)
whose position
$S_0$ is determined by equating these two fluxes, $F_1=F_2$,
\begin{equation}
\frac{ V_0 p_{on} m_f}{p_{off}}
=p_{end}l_0M_0+[p_{off}M_0-p_{on}m_f](l/2-S_0)
\end{equation}
This yields the following expression for the interface position
\begin{equation}
S_0=\frac{l}{2}
+\frac{p_{end}l_0M_0-V_0p_{on}p_{off}^{-1}m_f}{p_{off}M_0-p_{on}m_f}
\end{equation}

Obviously, $S_0$ grows with $p_{end}$, and at
$p_{end}=V_0p_{on}m_f[p_{off}l_0 M_0]^{-1}$, we obtain $S_0=l/2$,
i.e. the solid phase disappears.

Thus, the bound motor density $M_b$ is approximately described by the
following step function
\begin{equation}
M_b(S) \approx M_e + (M_0-M_e) \Theta(S-S_0) \label{step}
\end{equation}

The inhomogeneous distribution
of bound motors directly leads to the anisotropy of the
collision rate. The coordinates along the microtubules
$S_{1,2}$ are related to the positions ${\bf r}_{1,2} $ of the
center of microtubules as follows
\begin{equation}
S_{1,2} = {\bf n}_{1,2} \cdot ({\bf r} -{\bf r}_{1,2} )\label{s12}
\end{equation}
The collision rate is proportional to the sum of the bound motor
concentration times the cross-linking rate $g_0$,
\begin{equation}
g=g_0 [M_b(S_1^0)+M_b(S_2^0)] \label{kernel1}
\end{equation}
where $S_{1,2}^0$ are the values of $S_{1,2}$ at the intersection
point. Excluding ${\bf r}$ from Eq. (\ref{s12}) one obtains
\begin{eqnarray}
S_1^0&=&\frac{({\bf r_1} -{\bf r}_{2} )( {\bf n}_2-{\bf n}_1 ({\bf
n}_1
{\bf n}_2) ) }{1-({\bf n}_1 {\bf n}_2)^2} \nonumber \\
S_2^0&=&-\frac{({\bf r_1} -{\bf r}_{2} )( {\bf n}_1-{\bf n}_2
({\bf n}_1 {\bf n}_2) ) }{1-({\bf n}_1 {\bf n}_2)^2} \label{s120}
\end{eqnarray}
Non-uniform $M_b(S)$ produce anisotropy: $ g({\bf r}_1,{\bf
n}_1,{\bf r}_2,{\bf n}_2)\ne g({\bf r}_1,{\bf n}_2,{\bf
r}_2,{\bf n}_1)$. However, the collision rate Eq. (\ref{kernel1})
with the step-wise expressions for $S_{1,2}^0$ (\ref{step}) is awkward and impractical
for further calculations. In the subsequent section we will not
use the exact form (\ref{W1}) with Eq.(\ref{kernel1}) as a kernel in the master equation,
but replace it with a more simple
form which nevertheless retains the main features of
(\ref{W1}),(\ref{kernel1}), namely, localization and anisotropy,
\begin{equation}
W \approx   W_0\left( {\bf r_1} - {\bf r_2}\right) \left( 1- \beta
l^{-1}( {\bf r} _{1}- {\bf r} _{2}) ({\bf n} _{1} - {\bf n} _{2})
\right)  \label{W5}
\end{equation}
where the symmetric part of the kernel $W_0$ is of the Gaussian form
\begin{equation}
W_0({\bf r})=\frac{g_0}{\pi b^2} \exp[-{\bf r}^2/b^2]
\label{kernel_sym}
\end{equation}
with
the spatial scale $b\approx l/2$. The dimensionless parameter $\beta$
characterized the collision rate anisotropy.
The interaction kernel in this from was
proposed by us on the symmetry grounds in Ref. \cite{at_short}.

While the form (\ref{W5}) cannot be rigorously derived from (\ref{W1}),(\ref{kernel1}),
the anisotropy coefficient $\beta$ as a function of kinetic parameters
can be estimated from the expression (\ref{step}). First of all,
we approximate the step function in Eq. (\ref{step}) by the linear function
\begin{equation}
M_b(S) \approx \bar M + \bar \alpha S \label{step1}
\end{equation}
where the mean density $\bar M$ and mean slope $\bar \alpha$ are
calculated by the least mean square method using Eq. (\ref{step}),
\begin{eqnarray}
\bar M &=& \frac{1}{l} \int^{l/2} _{-l/2 } M_b dS =
\frac{M_0+M_e}{2}+ (M_0-M_e) \frac{S_0}{l} \nonumber \\
\bar \alpha &=&\frac{12}{l^3} \int^{l/2} _{-l/2 } M_b S dS= 6
\frac{M_0-M_e}{l^3} \left(\frac{l^2}{4}- S_0^2\right)
\label{slope}
\end{eqnarray}
To evaluate the effective kernel we  substitute Eq. (\ref{step1})
into Eq. (\ref{kernel1}) and using relations (\ref{s120}),
obtain
\begin{eqnarray}
g &\approx& g_0  \left[2\bar M  + \bar \alpha
(S_1^0+S_2^0)\right] \nonumber \\
&=&g_0 \left(2\bar M- \bar \alpha \frac{
({\bf r}_1 -{\bf r}_2) ({\bf n} _{1} - {\bf n} _{2})}{ 1-{\bf
n}_1 {\bf n}_2} \right)
 \label{sim_kern}
\end{eqnarray}
As one sees from Eq. (\ref{sim_kern}), it coincides with
phenomenological kernel Eq. (\ref{W5}) up to the factor
$(1-{\bf n}_1 {\bf n}_2)$ in denominator. The  value of the
dimensionless  kernel anisotropy is then
\begin{equation}
\beta= \frac{\bar \alpha l}{ 2\bar M} \label{anis1}
\end{equation}
Assuming that the density in the solid phase $M_0$ is much larger
than the density in the gas phase $M_e$, $\mu = M_0/M_e \gg 1 $,
we obtain the following estimate for the anisotropy $\beta$ for
$p_{end} \to 0$:
\begin{equation}
\beta \approx 3 \left(\frac{1}{2}-\frac{S_0}{l}\right)=3\frac{V_0-
p_{end} \mu l_0}{l p_{off} \mu}.
\end{equation}
The anisotropy is maximum for $p_{end}=0$, ($\beta \approx
V_0/p_{off}l \mu$), decreases with the increase of $p_{end}$  and
vanishes (at this approximation) at $p_{end}= V_0/\mu l_0$.

We can estimate the parameter $\beta$ for different type of motors
using the data from Ref. \cite{nedelec97,surrey01}. The parameters
for kinesin and NCD are: $V=1 $ $\mu/sec$, $p_{on}=20$ $sec^{-1}$,
$p_{off} =0.5$ sec$^{-1}$, and $p_{end}=70$ $sec^{-1}$ for kinesin
and $p_{end}=2.5$ $\sec^{-1}$ for NCd. Projected (two-dimensional)
density of free motors $m_f$ in Refs. \cite{nedelec97,surrey01}
was taken $m_f=0.05-2$ $\mu m^{-2}$. For the linear density of bound
motors we
obtain $M_e=p_{on} m_f d_0 /p_{off}$ and $M_0=1/l_0$, where $d_0
\approx 2 l_0$ is the diameter of microtubule. For parameter $\mu$
obtain   $\mu = M_0/M_e=p_{off} /(d_0 l_0  p_{on} m_f)\approx
10..500 \gg 1 $. Thus, for NCD-like motors when $l_0 \mu p_{end}
\ll 1$ we obtain the anisotropy parameter $\beta$ (depending on
the density ratios $\mu$)
\begin{equation}
\beta_{NCD} \approx \frac{3 V_0 }{l p_{off} \mu} \approx
10^{-3}.. 10^{-1} \label{beta_ncd}
\end{equation}
It follows from Eq. (\ref{beta_ncd}) that the kernel anisotropy increases
with the increase of the concentration of free motors.
Correspondingly, since for kinesin parameters
$p_{end}>V_0p_{on}m_f[p_{off}l_0 M_0]^{-1}$ and no solid phase is
formed, the anisotropy coefficient $\beta$ is essentially zero.

In this Section we considered only one mechanism contributing to
the kernel anisotropy: inhomogeneous distribution of bound motors.
Possibly there are other mechanisms affecting the anisotropy of
the interaction, for example,  finite bending rigidity of the
microtubules may also contribute to both the collision rules and the
collision rate.  However we leave this interesting topic to further
studies.

\section{Spatial localization of tubule-tubule interaction}
\label{sec2}
 To describe the {\em  spatial localization} of tubule-tubule
interaction we introduce the probability distribution $P(\bf r,
\phi, t)$ to find a rod with orientation $\phi$ at location ${\bf
r}$ at time $t$. The master equation for $P({\bf r}, \phi, t)$ can
be written as
\begin{widetext}
\begin{eqnarray}
&&\partial_t P({\bf r},\phi)=
\partial^2_\phi P({\bf r},\phi) + \partial_i D_{ij}
\partial_j P({\bf r}, \phi)
 + \int\int d{\bf r_1}d{\bf r}_2\int_{-\phi_0}^{\phi_0}dw \left[
W({\bf r}_1,{\bf r}_2,\phi+w/2,\phi-w/2) \right.
\nonumber\\
&& \times P({\bf r}_1,\phi+w/2)P({\bf
r}_2,\phi-w/2)\delta\left({{\bf r}_1+{\bf r}_2\over 2}-{\bf
r}\right)  \left.- W({\bf r}_1,{\bf r}_2,\phi,\phi-w)P({\bf
r_2},\phi)P({\bf r}_1,\phi-w)\delta\left({\bf r}_2 -{\bf
r}\right)\right] \label{master3}
\end{eqnarray}
\end{widetext}
where we performed the same rescaling as in Eq.(\ref{master2c})
using $g_0$ instead of $g$ and dropped argument $t$ for brevity.
We normalized the probability $P \to P/l^2$.

 Unlike the Maxwell model equation
(\ref{master2c}), Eq. (\ref{master3}) contains two diffusion terms
(translational and angular), and the motor-mediated tubule-tubule
collision integral now contains an interaction kernel which depends on
the relative tubule positions and orientations.
The angular diffusion coefficient $D_r$ and the
translational diffusion tensor $D_{ij}$ are known from the polymer physics
\cite{doi}:
\begin{eqnarray}
&&D_{ij}=\frac{1}{D_r} \left(D_\parallel n_i n_j + D_\perp
(\delta_{ij}-n_i n_j)\right)\nonumber\\
&&D_\parallel = \frac {k_B T}{\xi_\parallel}, \nonumber \\
&&D_\perp = \frac {k_B T}{\xi_\perp}, \nonumber \\
&&D_r=\frac {4 k_BT}{\xi_r} \label{diff}
\end{eqnarray}
where $\xi_\parallel, \xi_\perp, \xi_r $ are corresponding viscous
drag coefficients. For rigid rod-like molecules,
\begin{equation}
 \xi_\parallel= \frac{2 \pi
\eta_s l } {\log (l/d)} ; \;  \xi_\perp= 2 \xi_\parallel; \; \xi_r
\approx \frac{\pi \eta_s l^3}{ 3 \log (l/d)} \label{drug}
\end{equation}
where $\eta_s$ is shear viscosity \cite{doi}. Since we scaled time
by the rotational diffusion time $t \to D_r t$, in new variables
the translational diffusions  assume a very simple form: $
D_\parallel= 1/24, D_\perp=1/48 $, see Eq. (\ref{diff}). Note that
the drag coefficients are slightly modified for thin films and
membranes, see e.g \cite{levine}.

The last term  of Eq.(\ref{master3}) describes motors-mediated
interaction of rods. While the results of the previous section
indicate that the angular ``inelasticity coefficient'' $\gamma$ is
less than 1/2, we postulate here that after the interaction, the
two rods acquire the same orientation, $\phi=(\phi_1+\phi_2)$ and
the same spatial location in the middle of their original
locations, ${\bf r}=({\bf r}_1+{\bf r}_2)/2$. These assumptions
are made to simplify the calculations and final equations, however
generalization to arbitrary collision inelasticity is
straightforward (see Appendix \ref{app1}).

\subsection{Continuum equations}
\label{cont}

After integration over the $\delta$ functions Eq. (\ref{master3})
assumes the form
\begin{eqnarray}
&&\frac{\partial P}{\partial t}= \frac{\partial^2P}{\partial
\phi^2} +\partial_i D_{ij}
\partial_j P + Z_0+\beta Z_1 \label{master4a}
\end{eqnarray}
where nonlinear terms
\begin{widetext}
\begin{eqnarray}
&&Z_0=\int d{\bf r_1} \int_{-\phi_0}^{\phi_0}dw \left[ 2 W_0(2
({\bf r}_1- {\bf r})) P({\bf r}_1,\phi+w\gamma,t)P(2 {\bf r}-{\bf
r}_1,\phi+w(\gamma-1),t) \right.\nonumber\\
&& \left. - W_0({\bf r}_1-{\bf r})P({\bf r},\phi,t)P({\bf
r}_1,\phi-w,t)\right] \label{master4b}
\end{eqnarray}
and
\begin{eqnarray}
&&Z_1=\int d{\bf r_1} \int_{-\phi_0}^{\phi_0}dw \left[ 2
W_0(2 ({\bf r}_1- {\bf r}))({\bf r}_1-{\bf r})\cdot ({\bf
n}_1-{\bf n}) P({\bf r}_1,\phi+w\gamma,t)P(2 {\bf r}-{\bf
r}_1,\phi+w(\gamma-1),t)\right.
\nonumber\\
&& \left.- W_0({\bf r}_1-{\bf r})({\bf r}_1-{\bf r})\cdot ({\bf
n}_1-{\bf n})P({\bf r},\phi,t)P({\bf r}_1,\phi-w,t)\right]
\label{master4c}
\end{eqnarray}
\end{widetext}
are generated by the collision
integral in Eq. (\ref{master3}).

To proceed, we again perform the Fourier expansion  of the probability
distributions in $\phi$ and truncate the series at
$|n|>2$. Now $2 \pi P_0$ gives the local number density $\rho({\bf r},
t)$, and $P_{\pm1}$ the local orientation $\btau({\bf r},t)$.
The
integration of the diffusion term in Eq.(\ref{master4a}) generates
linear terms, and
the nonlinear terms $Z_0, Z_1$ (see Appendix \ref{app2}) produce
nonlinear terms in the corresponding equations for $\rho, \btau$.

After  rescaling space by $l$, and introducing dimensionless
parameters $ B=b/l$ characterizing the width of the interaction
kernel and $H=\beta  B^2 $ characterizing normalized strength of
anisotropy of interaction, we arrive at the set of equations for
coarse-grained local density $\rho$ and orientation $\btau$
\begin{widetext}
\begin{eqnarray}
\partial_t \rho &=&   \nabla^2 \left[ \frac{\rho}{32}
-{B^2 \rho^2 \over 16 }\right]  -{\pi B^2H\over 16} \left[ 3
\nabla \cdot  \left( \btau  \nabla^2\rho - \rho \nabla^2 \btau
\right)+
 2 \partial_i \left(
\partial_j \rho \partial_j \tau_i   - \partial_i \rho \partial_j
\tau_j \right) \right] -\frac{7 \rho_0 B^4}{256} \nabla^4 \rho
\label{rho_1}
\\
\partial_t{\btau} &=&
\frac{5}{192} \nabla^2 \btau + \frac{1}{96}\nabla (\nabla \cdot
\tau ) +\epsilon\btau -A_0|\btau|^2\btau
 - H\left[\frac{\nabla\rho^2}{16 \pi}- \left(\pi-{8\over
3}\right) \btau(\nabla\cdot\btau)- {8\over 3} (\btau \nabla) \btau
\right] +\frac{B^2 \rho_0}{4 \pi}\nabla^2 \btau   \label{tau_1}
\end{eqnarray}
\end{widetext}
These equations generalize Eq.(\ref{Pk2c}) for the case of
spatially localized coupling. For simplicity the last two terms in
Eqs. (\ref{rho_1}),(\ref{tau_1}) have been linearized near the
mean density $\rho_0=\langle \rho \rangle$, otherwise more
complicated expressions given in Appendix \ref{app2} are needed.
This approximation is justified by the fact that in the relevant
range of parameters $B, H$ the density variations are small
compared to the variations of the orientation $\btau$. The last
term in Eq. (\ref{rho_1}) regularizes the short-wave instability
when the diffusion term changes sign for $\rho_0>\rho_b=1/4 B^2$.
This instability leads to strong density variations associated
with formation of dense microtubule bundles (see Figs.
\ref{fig:bound}, \ref{fig:bundle} below)  which is also observed
experimentally for large density of molecular motors.

\subsection{Stability of Asters and Vortices}
\label{sec3}

\begin{figure}[ptb]
\includegraphics[width=3.2in,angle=0]{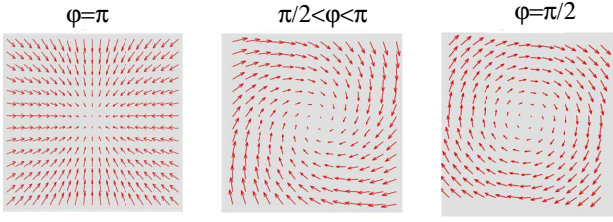}
\caption{Schematic representation of orientation fields $\btau$
for three different values of $\varphi$: aster ($\varphi=\pi$);
generic vortex ($\pi/2<\varphi<\pi$) and ideal vortex
($\varphi=\pm \pi/2$).  } \label{fig:three}
\end{figure}

If $B^2H\ll 1$, the density modulations are rather small,  and Eq.
(\ref{tau_1}) for orientation $\btau$ decouples from Eq.
(\ref{rho_1}). It is convenient to rewrite Eq. (\ref{tau_1}) for
complex variable $\psi=\tau_x+i \tau_y$:
\begin{eqnarray}
&&\partial_t \psi= \epsilon \psi-A_0 |\psi|^2 \psi + D_1 \nabla^2
\psi
+ D_2 \bar \nabla ^2 \psi^* \nonumber \\
&&+H\left(\left(\pi-\frac{8}{3} \right) \psi \mbox{Re}\bar \nabla
\psi^* + \frac{8}{3}  (\mbox{Re}\psi^* \bar \nabla )  \psi
 \right) \label{psi1}
\end{eqnarray}
where operator $\bar \nabla = \partial_x +i \partial_y $, $
D_1=1/32+ \rho_0 B^2/4 \pi, D_2=1/192$. Eq. (\ref{psi1}) is
similar to generalized Ginzburg-Landau equation known in the
context of superconductivity, superfluidity, nonlinear optics, and
pattern formation, see e.g. \cite{ArKr}. Let us focus on the
structure and the dynamics of radially-symmetric solutions of
(\ref{psi1}) which can be sought in polar coordinates $r, \theta$
in the following generic form:
\begin{equation}
\psi= \sqrt{A_0/\epsilon} A(r) \exp ( i \theta)\label{psi2}
\end{equation}
where the complex amplitude $A(r)=\Phi(r) \exp[i \varphi(r)]$, and
the phase $\varphi(r)$ is a real function. The solution
$\varphi(r)=0,\pi$ corresponds to asters, and the solution with
$\varphi(r)\ne 0$ describes vortices, see Fig. \ref{fig:three}.
Transitions between asters and vortices can be examined in the
framework of a one-dimensional problem for the complex variable
$A(r)$,
\begin{eqnarray}
&&\partial_t A = D_1 \Delta_r A + D_2 \Delta_r A^* + \left( 1-| A|
^2 \right) A \nonumber
\\
&&+ H \left ( a_1 A \mbox{Re} \nabla_r A + a_2
\partial_r A  \mbox{Re} A+ \frac{ i a_2 A \mbox{Im} A}{r} \right)
\label{rad2}\end{eqnarray} with the following differential
operators
\begin{equation}
\Delta_r=\partial_r^2+r^{-1}
\partial_r-r^{-2}; \nabla_r=\partial_r+r^{-1}
\end{equation}
and parameters
\begin{eqnarray*}
a_1&=&(\pi-8/3)/\sqrt{A_0}\approx 0.321, a_2=8 /3
\sqrt{A_0}\approx 1.81
\end{eqnarray*}
Here  we rescaled time $t \to t/\epsilon$ and space variable  by
$r \to r/\sqrt{\epsilon}$.

The aster and vortex solutions obtained by numerical integration
of Eq. (\ref{rad2}) for certain parameter values are shown in Fig.
\ref{fig:vort_ast}. Vortices exist only for small values of $H$
and give way to asters for larger $H$ or larger $\rho$. For $H=0$,
Eq.(\ref{rad2}) reduces to a form which was studied in
\cite{at03}. It was shown in \cite{at03} that the term $\Delta_r
A^*$ favors vortex solution $(\varphi= \pm \pi/2)$. In contrast,
terms proportional to $H$ select asters with $\varphi=\pi$ (aster
for $\varphi=0$ is unstable for the anisotropy parameter $H>0$).
Note that $\varphi=\pi$ corresponds to asters with the direction
of arrows towards the center, as it is shown in Fig.
\ref{fig:three}. Since we associated the direction of the vector
$\tau$ with the direction of motion of molecular motors along the
microtubules, the aster with $\varphi=\pi$ corresponds to the
experimental situation: motor moves towards the center of the
aster. Increasing $H$ leads to gradual reduction of $\varphi$, and
at a finite $H_0(\rho_0)$, $\phi(r)=\pi$, i.e. the transition from
vortices to asters occurs. For $0<H<H_0$, the vortex solution has
a non-trivial structure. As seen in Fig. \ref{fig:vort_ast}, the
phase $\varphi \to \pi$ for $r \to \infty$, i.e.  vortices and
asters become indistinguishable far away from the core.

The transitions between asters and vortices can be studied in the
framework of linearized Eq. (\ref{rad2}). For this purpose the
solution to Eq. (\ref{rad2}) can be sought in the form
\begin{equation}
A(r)=\Phi(r)+i w(r) \exp(\lambda t)
\end{equation}
where small real perturbation $w$ obeys a linear equation
$\hat L=\lambda w$ with operator
\begin{equation}
\hat L\equiv \bar D \Delta_r  + \left( 1-\Phi ^2 + a_1 H  \nabla_r
 \Phi \right)  +a_2 H \Phi \nabla_r \label{rad3}\end{equation}
($\bar D=D_1-D_2$) with zero boundary conditions at $r=0,\ \infty$.
This eigenvalue problem can be solved by the matching-shooting method. A
positive eigenvalue $\lambda$
corresponds to the emergence of a non-zero phase $\phi(r)$, i.e.  a vortex.

\begin{figure}[ptb]
\hspace{-1.5cm}
\includegraphics[width=1.8in,angle=-90]{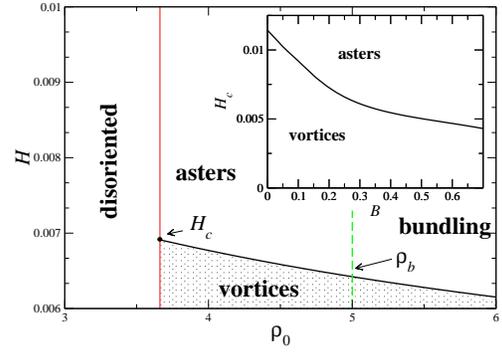}
\caption{Phase boundaries obtained form the linear stability
analysis of the aster solution  for $B^2=0.05$, dashed line shows
bundling instability limit $\rho_0=\rho_b=5$. Inset: Position of
critical point $H_c$ vs $B$ at $\rho_0=4.5$.} \label{fig:bound}
\end{figure}

\begin{figure}[ptb]
\hspace{-1cm}
\includegraphics[width=1.8in,angle=-90]{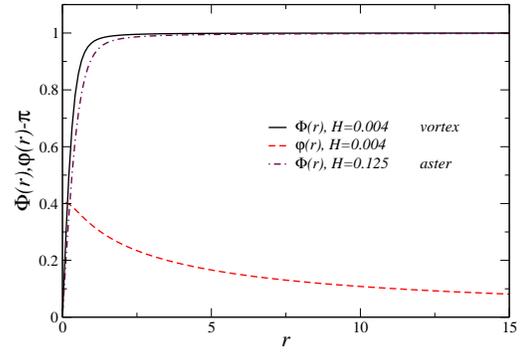}
\caption{Stationary  vortex and aster solutions $\tau_x+i
\tau_y=\Phi(r) \exp[i \theta+ i \varphi(r)]$ to Eq. (\ref{rad2}),
for $\rho_0=4, B^2=0.05$.} \label{fig:vort_ast}
\end{figure}

\begin{figure}[ptb]
\includegraphics[width=3.in,angle=0]{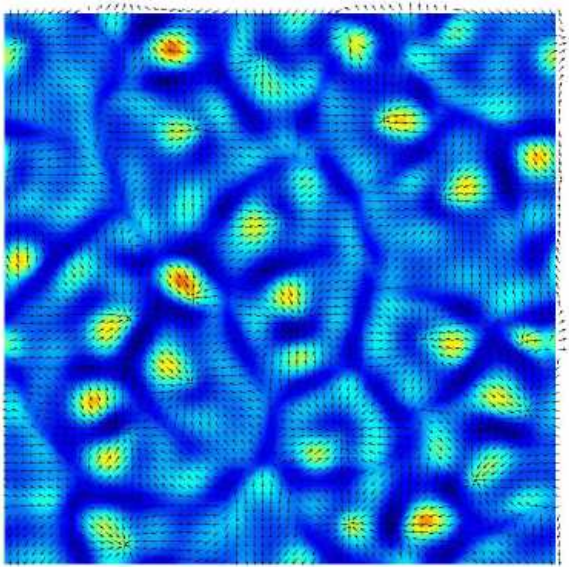}
\caption{Composite image of the density (colors) and orientation
(arrows)  fields in the regime of density instability. Density
changes from $\rho_{max} \approx 10$ (red) to $\rho_{min} \approx
4$ (dark blue).  Parameters:  $B^2=0.05, \rho_0=6$, $H=0.125$,
domain of integration $80\times 80$ units.} \label{fig:bundle}
\end{figure}

The resulting phase diagram of the continuum model (\ref{rho_1}),
(\ref{tau_1}) is shown in Fig. \ref{fig:bound}. The solid line
$H_0(\rho_0)$ separating vortices from asters is obtained from the
solution of the linearized Eq. (\ref{rad2}) by tracking the most
unstable eigenvalue $\lambda$ of the aster. The dashed line
corresponds to the onset of the orientation instability,
$\rho_0=\rho_c$. The lines meet at the critical point
$H_c=H_0(\rho_c)$ above which vortices are unstable for arbitrary
small $\epsilon>0$. The phase diagram is qualitatively consistent
with experiments, see  Ref. \cite{surrey01}: for low value of
kernel anisotropy $H<H_c$ (which according to our estimates in
Sec. \ref{sec:kern} correspond to kinesin-like motors with very
small anisotropy value $\beta$) increase of the density $\rho_0$
first leads to formation of vortices and then asters. For  $H>H_c$
(which apparently correspond to Ncd motors with large anisotropy)
only asters are observed. For large density values in addition to
orientation instability one observes density instability
$\rho_0>\rho_b= 1/4 B^2$ produced by Eq. (\ref{rho_1}) when the
diffusion term changes sign. Numerical studies of the full system
(\ref{rho_1}), (\ref{tau_1}) indeed indicate formation of extended
domains of high density not associated with the asters, see Fig.
\ref{fig:bundle}. While our model (\ref{rho_1}), (\ref{tau_1})
yields the density instability and bundle formation in accordance
with experiment, we anticipate only qualitative agreement in this
regime because the model itself is derived in the low density
limit when only binary interactions are included.

\subsection{Interaction of asters and vortices}
For $H \ne 0$ well-separated vortices and asters exhibit
exponentially weak interaction. For asters it follows from the
fact that $\hat L$ is not a self-adjoint operator. To investigate
the interaction between asters we need to examine the asymptotic
null-space of the adjoint operator $\hat L^\dagger$
for $r \to \infty$  (see
for details \cite{ArKr}). After simple algebra we obtain that for
$r \to \infty$ the adjoint operator is of the form
\begin{equation}
\hat L^\dagger \equiv \bar D \partial_r^2    -a_2 H  \partial_r
\label{rad4}\end{equation} Substituting the solution to Eq.
(\ref{rad4}) in the form $ w^\dagger \sim \exp [p r ] $ we obtain
that there are two solutions: $p=0$ which describes the neutral
translation mode, and the non-trivial solution
\begin{equation}
p=a_2 H/\bar D \label{s_leng}
\end{equation}
Since due to interaction the well-separated asters produce only
small perturbations to their shape, these perturbations    can be
treated in linear approximation, and
 the exponent (\ref{s_leng})  characterizes  asymptotic screening of
the interaction between the asters analogous to the interaction of
spiral waves in the Ginzburg-Landau equation, see Ref. \cite{ArKr}
for details of analysis. Thus we obtain that perturbations
produced due to interaction of well-separated asters  decay as $w
\sim \exp[-r/L_0]$  with the screening length $L =1/p$, or in
original units $L_0 = \bar D /a_2H\sqrt{\epsilon}=\bar D
/a_2H\sqrt{\rho/\rho_c-1}$ (see for details \cite{ArKr}).
Screening length  $L_0$ diverges for $H \to 0$ and at the
threshold $\rho_0 \to \rho_c$. Similar analysis can be performed
for vortices.

\begin{figure}[ptb]
\includegraphics[width=1.68in,angle=0]{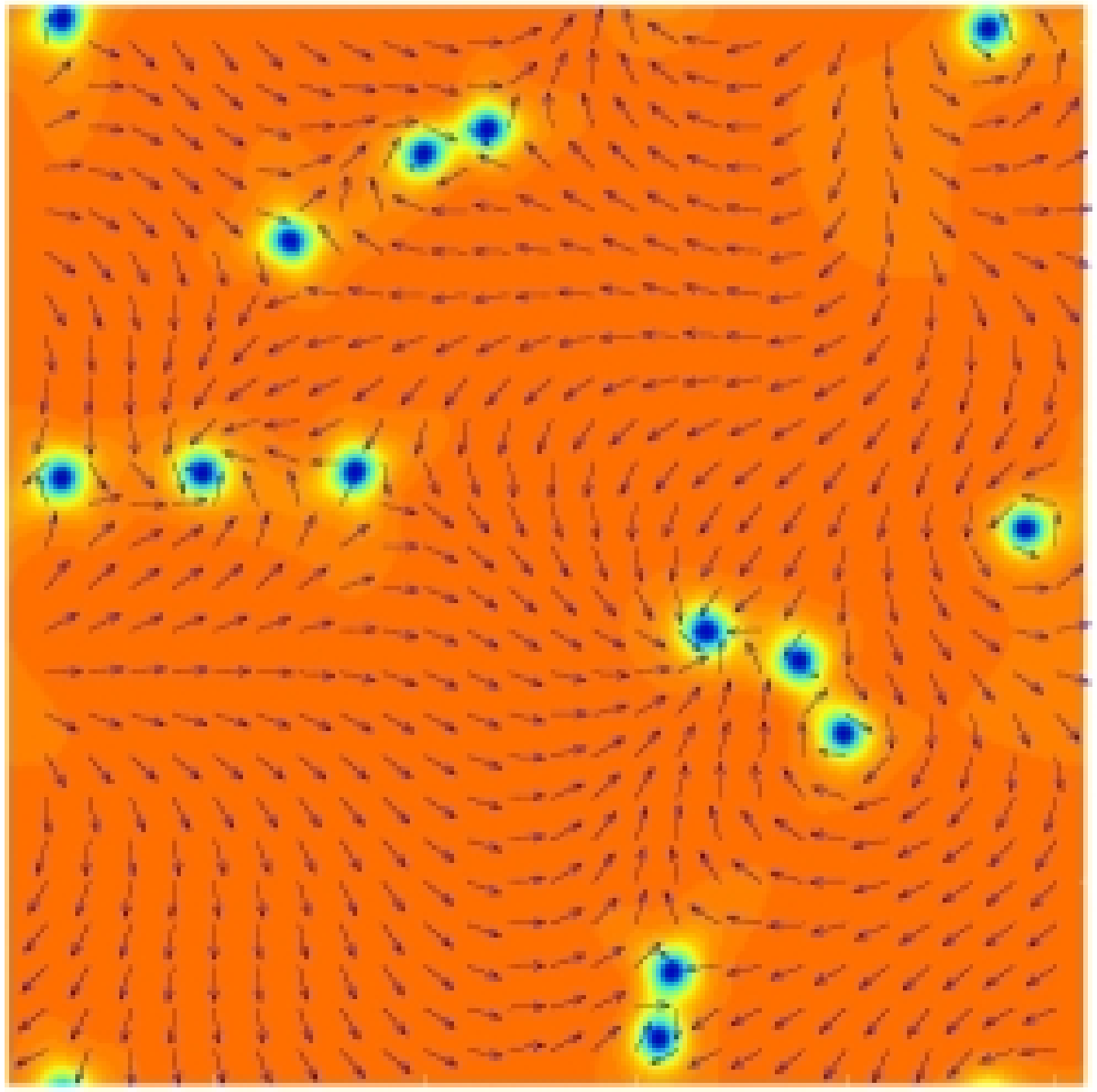}
\includegraphics[width=1.68in,angle=0]{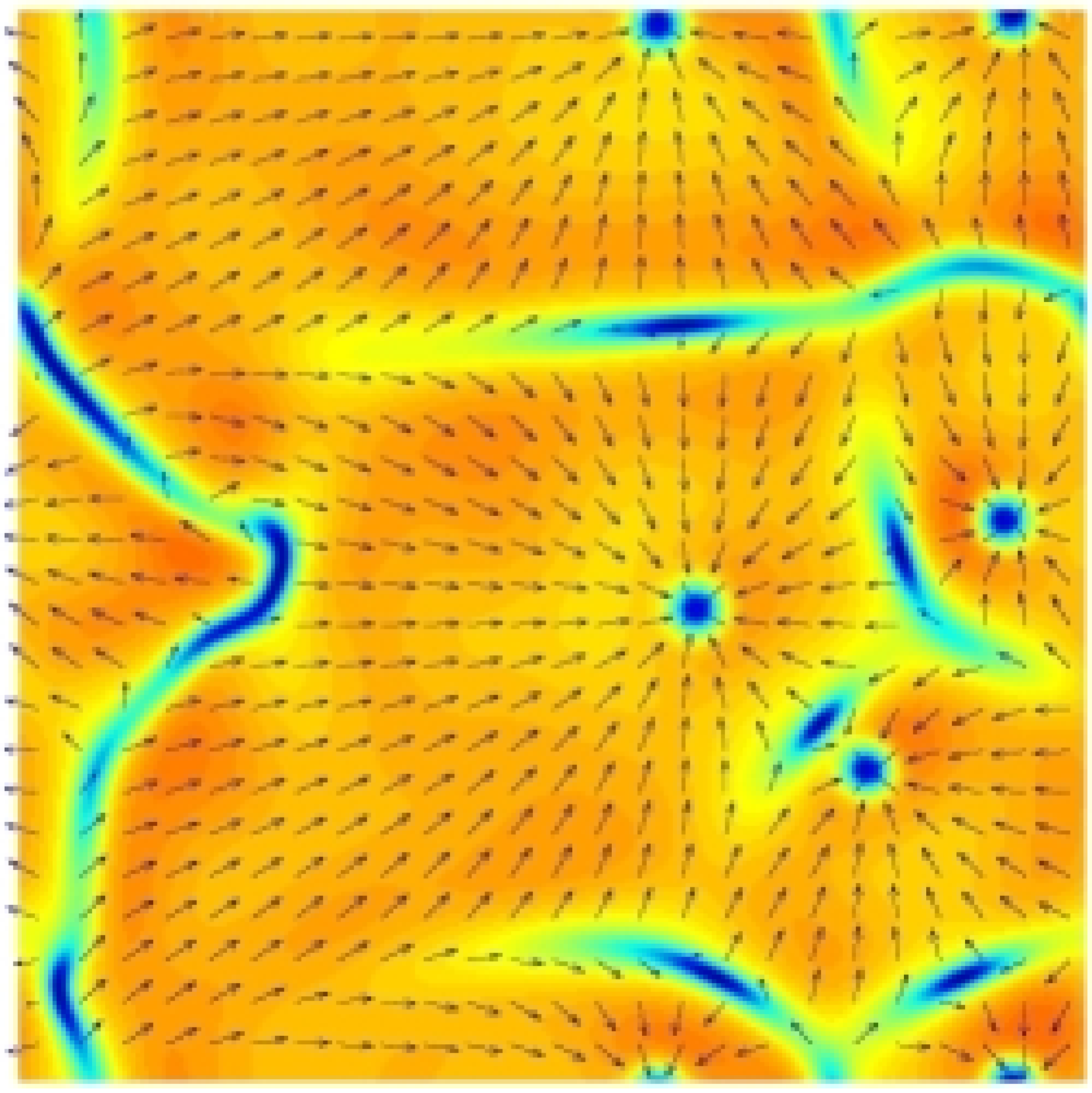}
\caption{Orientation  $\tau$ for vortices ($H=0.006$, left) and
asters ($H=0.125$, right) obtained from Eqs.
(\ref{rho_1}),(\ref{tau_1}). Color code indicates the intensity of
$|\tau|$ (red corresponds to maximum and blue to zero), $B^2=0.05,
\rho_0=4$, domain of integration $80\times 80$ units, time of
integration 1000 units. } \label{fig:av}
\end{figure}

\subsection{Numerical solution of full system}
 We also studied the full system (\ref{rho_1}),(\ref{tau_1})
numerically. Integration was performed in a two-dimensional square
domain with periodic boundary conditions by the quasi-spectral method.
For small $H$ we observed vortices and for larger $H$ asters, in
agreement with the above analysis.  Since vortices exist for
smaller values of $H$, their screening length
$L$  is larger than for the asters because $L \sim 1/H$, see Eq.
(\ref{s_leng}). Thus, the vortices interact stronger and are more
keen to annihilate than asters. As seen in Fig. \ref{fig:av},
asters have a unique orientation of the microtubules (here, towards
the center). Asters with the opposite orientation of $\tau$ are
unstable.

In large domains asters form a disordered network of cells with the
cell size of the order of $L_0$. Neighboring cells are separated
by the ``shock lines'' terminated by  saddle-type defects. The
pattern of asters resembles a ``frozen'' glass state of spirals observed
in the complex Ginburg-Landau model
\cite{ArKr,brito}. Starting from a random initial condition we
observed initial merging and annihilation of asters. Eventually,
annihilation slows down due to exponential weakening of the
interaction of asters. For the same integration time the number of
vortices is typically smaller than the number of aster due to the
fact that the screening length of asters is smaller.

\subsection{Drift instability of the asters}
\begin{figure}[ptb]
\includegraphics[width=3in,angle=0]{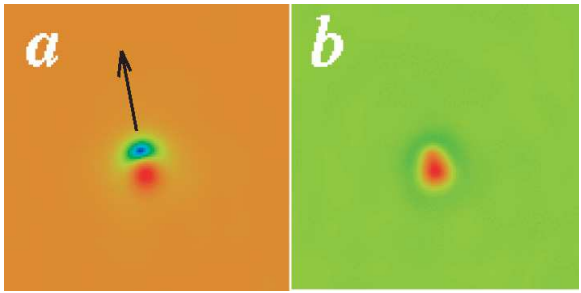}
\caption{Drift instability of asters for $H=0.125,B=0.06$,
$\rho_0=4$, size of the image  $40\times40$ at the moment of time
$t=300$. Left image shows $|\btau|$, right image shows $\rho$, arrow
indicates the direction of drift. Color code: blue corresponds to
zero, red corresponds to maximum.} \label{fig:acc}
\end{figure}

\begin{figure}[ptb]
\includegraphics[width=3.2in,angle=0]{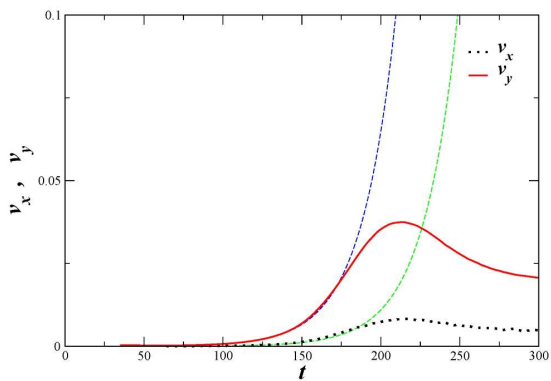}
\caption{Aster core velocity vs time for parameters of Fig.
\ref{fig:acc}. Solid line shows $v_y$, dotted line shows $v_x$,
dashed lines depict exponential fit $v\sim \exp(\lambda t)$ for
first 150 units of time.} \label{fig:exp}
\end{figure}

In experiments \cite{nedelec97,surrey01} asters
often are  not stationary:  they drift and coalesce.
Surprisingly, in our numerical investigations of Eqs.
(\ref{rho_1}), (\ref{tau_1}) we also observed that typically the
center of an aster is unstable and develops a spontaneous
acceleration instability, see Fig. \ref{fig:acc}. This instability
is reminiscent of the instability of the core of spiral waves in
the complex Ginzburg-Landau equation in a large dispersion limit,
see Ref. \cite{akw}. This instability associated in Ref.
\cite{akw} with the exponential growth of localized mode in the
form $w_1(r) \exp[i \theta]$, which is similar to the translation
mode and results in the displacement of the core. We have found
that the instability can be suppressed by increasing
the coefficients in front of the last term ($\sim \nabla^4 \rho$)
in Eq. (\ref{rho_1}). In this context it is possible that the
acceleration instability is just an artifact of the approximations
made in the course of derivation, such as binary character of interactions of
microtubules, small-gradient expansion, etc. Furthermore,
for high density of the microtubules
the prefactor in form of the cut-off term $\sim \nabla^4 \rho$ to
should grow rapidly  and dampen the instability. While at the moment
there is no unambigous experimental evidence of this instability
(e.g. the drift of asters could be also attributed  to gradients
of microtubules or motor distributions, effects of the boundaries
etc), we believe that this instability  can be found in
certain experimental conditions.
To study the drift instability we prepared by the initial
condition axisymmetric aster solution perturbed by the small
amplitude noise. In the course of motion the solution breaks the
axial symmetry,  typical structure of the moving aster is shown in
Figure \ref{fig:acc}. There is a small but noticeable (about 10\%)
increase of the density $\rho$ and the amplitude of orientation
$|\btau|$ behind the aster, for the immobile solution the position
of the zero of $\btau$ and maximum of $\rho$ coincides. The
instability accelerates collisions and coalescence of asters.
However the growth rate $\lambda$ of the instability appears to be
very small and aster solutions are well-preserved for a very long
time (several hundreds of dimensionless units). Fig. \ref{fig:exp}
shows the velocity of aster core vs time for the parameters of
Fig. \ref{fig:acc}. One clearly sees initial exponential growth of
the aster velocity.

\section{Effects of motors attached to the bottom plate}
\label{sec4}

In the previous Sections we considered microtubules interacting
with molecular motors freely floating in the solvent. However, in
{\em in vitro} experiments it is difficult to prevent attachment
of some fraction of motors to the bottom of the cell with one of
their two heads. The other (free) head of the  attached (absorbed)
molecular motor then may bind to a microtubule and push it in the
direction opposite its orientation. This effect was observed
experimentally in Ref. \cite{vale} (refereed to as microtubule
gliding assays).

The effect of attached motors can be easily incorporated in the
master equation
\begin{eqnarray}
&&\frac{\partial P}{\partial t}= \frac{\partial^2P}{\partial
\phi^2} +\partial_i D_{ij}\partial_j P +\alpha \nabla ({\bf n} P)
 + Z_0+Z_1 \label{master5a}
\end{eqnarray}
Here $\alpha$ is the fraction of the attached motors, and the term
${\bf n} \nabla P$ account for the transport of the microtubules
in the direction opposite to their orientation vector $\bf
n=[\cos\phi,\sin\phi]$. Terms $Z_{0,1}$ remain unchanged. It is
easy to check that the drift term will generate additional linear
terms $\alpha \pi \nabla  \btau$ in Eq. (\ref{rho_1}) and
$\alpha(4 \pi)^{-1} \nabla \rho$ in Eq.(\ref{tau_1}).

It is useful to write  Eq. (\ref{rho_1}) in the form of the mass
conservation law
\begin{equation}
\rho_t = - \nabla \bf J
\end{equation}
with the corresponding mass flux ${\bf J}$. The anisotropic part
of the kernel generates mass flux which is second order in the
gradients of  $\rho, \btau$. The the lowest-order term ${\bf J}
\sim \btau$  in the expression for flux is generated by the motors
attached to the substrate and is similar to that of self-propelled
particles, see e.g. \cite{at03}. In the situation considered in
Sec. \ref{cont} this term is prohibited by the momentum
conservation: the molecular motors produce only internal forces
which cannot displace the center of mass of the system. However
this is not the case when some of the motors are attached to the
substrate. In Ref. \cite{marchetti05} similar contribution to the
flux were attributed to the net displacement of the center of the
microtubule pair due to the anisotropy of the viscous drag
coefficient. However this pure hydrodynamic effect is probably
smaller than the advection produced by the motors absorbed at the
substrate.

\begin{figure}[ptb]
\includegraphics[width=3in,angle=0]{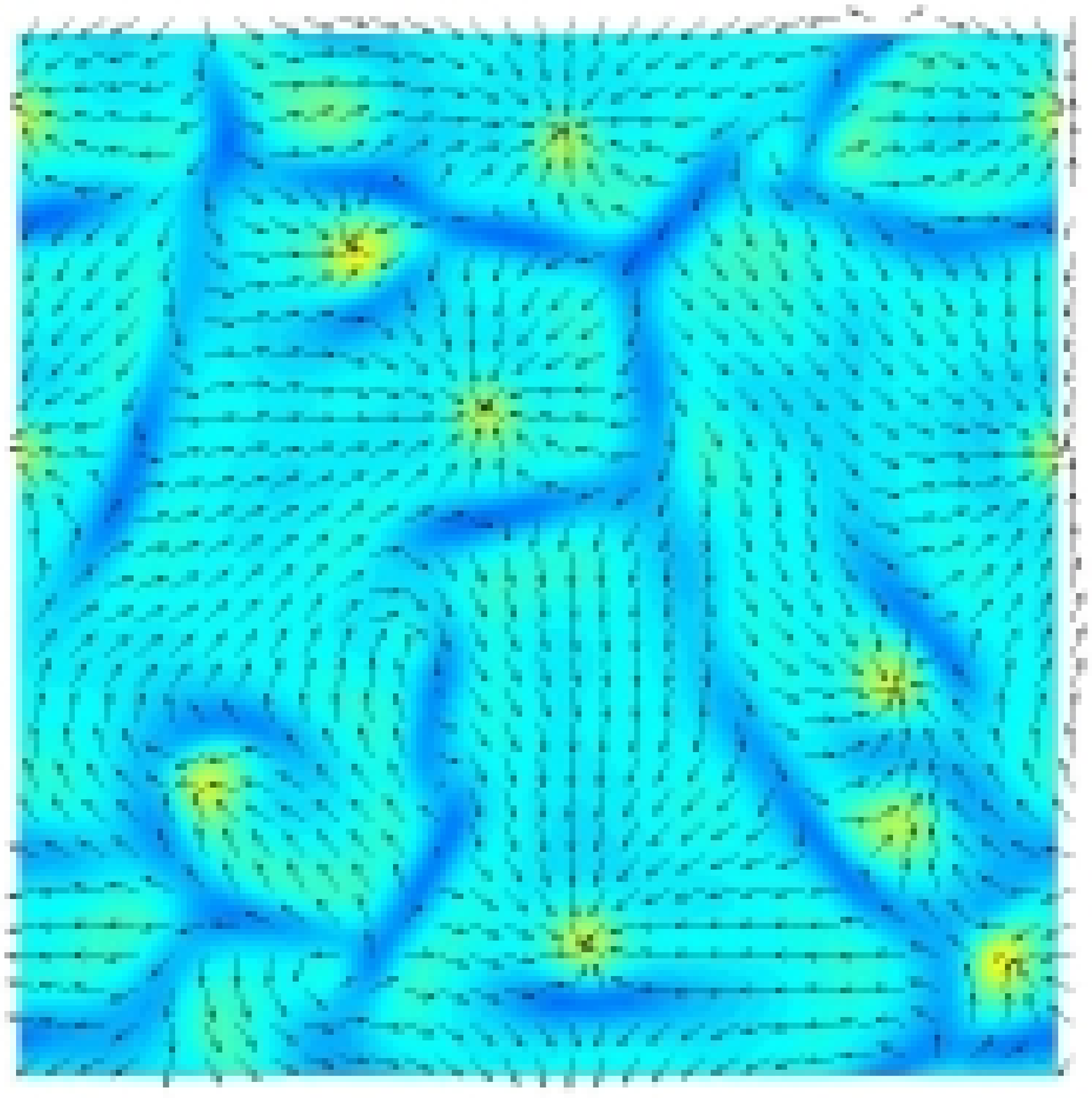}
\includegraphics[width=3in,angle=0]{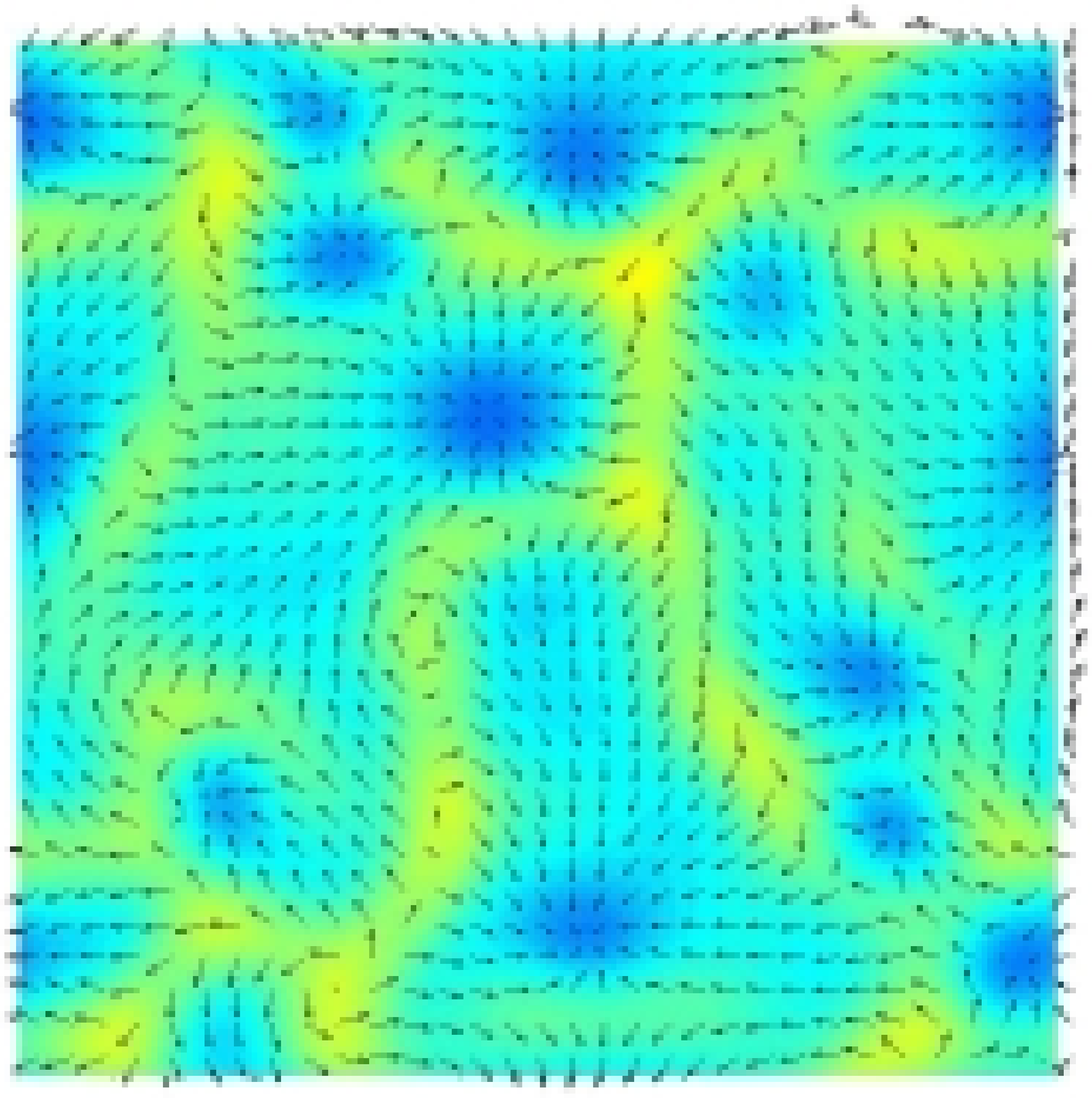}
\caption{Comparison between density distributions (colors) without
(upper image) and with drift term ($\alpha=0.004$). Arrows show
corresponding orientation. overall change in the density about
5\%, $B^2=0.05, \rho_0=4$, $H=0.1$, domain of integration
$80\times 80$.} \label{fig:drift}
\end{figure}

While the fraction of absorbed motors $\alpha$ might be small, it
still can produce a considerable effect because it generates the
lowest-order transport term in Eqs. (\ref{rho_1}),(\ref{tau_1}).
Numerical studies of Eqs. (\ref{rho_1}),(\ref{tau_1}) with
additional $\alpha$-terms reveal that the qualitative features are
not very sensitive to the presence of these terms for small
$\alpha \ll 1$, as long as the diffusive transport in the equation
for the density $\rho$ dominates advection. However, for moderate
$\alpha $ we observed that the aster and vortices become even less
localized, see Fig. \ref{fig:drift}. This delocalization is due to
the fact that the absorbed motors advect the mocrotubules in the
direction opposite their orientation. Consequently, these motors
move the tubules {\it from the asters} and make a small depression
of density for $\alpha \ne 0$ contrary to the density peak for
$\alpha =0$, compare images on Fig. \ref{fig:drift}. Similar
results are also obtained for vortices. Remarkably, the
suppression of  density of microtubules in the core of vortex is
observed experimentally, see Fig. 2a in Ref. \cite{surrey01}.

The absorbed motors, resulting  in the displacement of the center
of mass of the microtubules system, may  explain rotation of
vortices absent in our previous analysis. Indeed, since these
motors generate net motion of individual microtubules with the
velocity $\sim \alpha$, they can support rotating configurations
similar to that observed in the system of vibrated rods
\cite{at03}. Obviously no rotation anticipated for asters due to
pure radial orientation of microtubules: the forces induced by
motors attached to substrate will be compensated by ``pressure''
gradient due to redistribution of density of microtubules. In
contrary, the rotation is present for vortices. Far away from the
core the distinction between vortex and aster disappears, the
rotation is localized only at the core of the vortex where the
phase $\phi$ is different from $\pi$. Since the amplitude of
orientation vector $\tau$ grows almost linearly from the vortex
core and reaches asymptotic value $\tau_0 \approx
\sqrt{\epsilon/A_0}$ at the distance about $1-2$ dimensionless
units (see Fig. \ref{fig:vort_ast}), the rotation frequency of the
vortex core $\omega \approx \alpha \tau_0$. Indeed, rotation of
the vortex core was observed experimentally. For the parameters of
our numerical studies the frequency $\omega $ is very small due to
the smallness of $\alpha=0.004$, thus during the time of numerical
experiment ($<1000$ dimensionless units of time) the vortex core
turned only the fraction of full circle.

\section{Inhomogeneous distribution of motors}
\label{sec5}

In previous Sections we always assumed a homogeneous bulk distribution of
molecular motors (however we took into account local inhomogeneity of
bound motor concentration on the scale of a single tubule to account for
the collision rate anisotropy in Sec.\ref{sec:kern}).
This assumption was justified by the fact that the
diffusion of motors is about two orders of magnitude larger than of
microtubules. However experiments indicate that even despite this
strong diffusion, molecular
motors aggregate in the core regions of asters and vortices due to the
directed transport of motors by microtubules \cite{nedelec01}.

To describe the dynamics of the motor concentration we again invoke the
equations for free $m_f$ and bound $m_b$ motor concentrations, but unlike Sec.
\ref{sec:kern}, we will coarse-grain these distributions on the scale much
larger than the size of individual filament similar to Ref.
\cite{nedelec01} (see also
\cite{lee01,kim03,sank})
The populations $m_b,m_f$ obey the advection-diffusion equations
\cite{nedelec01} (compare with Sec. \ref{sec:kern})
\begin{eqnarray}
\partial_t m_f & =&  D \nabla^2 m_f - \rho ( p^{on} m_f- p^{off} m_b)
\nonumber \\
\partial_t m_b & =& - \zeta \nabla m_b \btau   + \rho ( p^{on} m_f- p^{off} m_b)
\label{motor1}
\end{eqnarray}
where $ p^{on}, p^{off}$ are the rates of binding/unbinding of
motor to the microtubules, $D,\zeta$ are diffusion/advection
coefficients accordingly.

If we assume that the distributions of $m_f,m_b$ are smooth and
the binding/unbinding rates are large,  then the r.h.s.  Eqs.
(\ref{motor1}) is dominated by the last term describing
binding/unbinding of the motors, leading to the local balance
relation  between $m_f, m_b$
\begin{equation}
p^{on} m_f \approx  p^{off} m_b \label{motor3}
\end{equation}
Then we can reduce system (\ref{motor1}) to a single equation for the
total motor density $m=m_f+m_b$:
\begin{equation}
\partial_t m =  D_0 \nabla^2 m -  \zeta_0 \nabla m \btau
\label{motor4}
\end{equation}
where $D_0=D p^{off} /(p^{off}+p^{on}) $ and $\zeta_0=\zeta p^{on}
/(p^{off}+p^{on}) $.

Accordingly, we need to modify the expression for the interaction
kernel Eq. (\ref{W5}) in order to include the effect of the
motors. The simplest way to include inhomogeneous motor density
into the kernel is the following,
\begin{eqnarray}
 W_m&=& m \left(\frac{{\bf r}_1+{\bf r}_2}{2}
\right) W({\bf r}_1-{\bf r}_2, \phi_1-\phi_2 )  \label{kern2}
\end{eqnarray}
where $W$ is given by Eq. (\ref{W5}). Taking the motor
concentration in the middle point $({\bf r}_1+{\bf r}_2)/2 $ is
necessary to preserve the mass  conservation law.
Repeating the calculations presented in the
previous sections one can derive equations similar to Eqs.
(\ref{rho_1}),(\ref{tau_1}) but with the motor density as an
independent field. However the resulting equations are very
cumbersome, especially for the transport term in Eq.
(\ref{rho_1}).

One can simplify the problem considerably utilizing again the fact
that the motor diffusion is high, and therefore, the
distribution of $m$ is smooth. Then, one can neglect the derivatives
of $m$ where it is appropriate, and the resulting equations assume
the form:

\begin{widetext}
\begin{eqnarray}
\partial_t \rho &=&   \nabla^2 \left[ \frac{\rho}{32}
-{m B^2 \rho^2 \over 16 }\right]+\alpha \pi \nabla \btau \nonumber
\\ &-&{\pi B^2H\over 16} \left[ 3 \nabla  m \left( \btau
\nabla^2\rho - \rho \nabla^2 \btau \right)+
 2 \partial_i m \left(
\partial_j \rho \partial_j \tau_i   - \partial_i \rho \partial_j
\tau_j \right) \right] -\frac{7 \rho_0  m_0 B^4}{256} \nabla^4
\rho \label{rho_2}
\\
\partial_t{\btau} &=&
\frac{5}{192} \nabla^2 \btau + \frac{1}{96}\nabla (\nabla \cdot
\tau )+\frac{\alpha}{4 \pi} \nabla \rho +\left( (4/\pi-1) m \rho -1) \right) \btau \nonumber \\
&-&A_0|\btau|^2\btau
 - H m \left[\frac{\nabla\rho^2}{16 \pi}- \left(\pi-{8\over
3}\right) \btau(\nabla\cdot\btau)- {8\over 3} (\btau \nabla) \btau
\right] +\frac{B^2 \rho_0 m_0 }{4 \pi}\nabla^2 \btau \label{tau_2}
\end{eqnarray}
\end{widetext}

Thus, motor density is included in the lowest order in gradient
expansion. We also included terms $\sim \alpha$ describing the
transport of microtubules by the absorbed motors. Again for
simplicity we replaced the motor density $m$ by its mean value
$m_0$ in the last terms in Eqs. (\ref{rho_2}),(\ref{tau_2}).

We carried out numerical studies of  Eqs.
(\ref{rho_2}),(\ref{tau_2}). The values of the parameters $D_0,
\xi_0$ can be estimated from the experimental conditions, in our
dimensionless units $D_0 \sim 1..5$ and $\xi_0 \sim 1$.

\begin{figure}[ptb]
\includegraphics[width=3in,angle=-90]{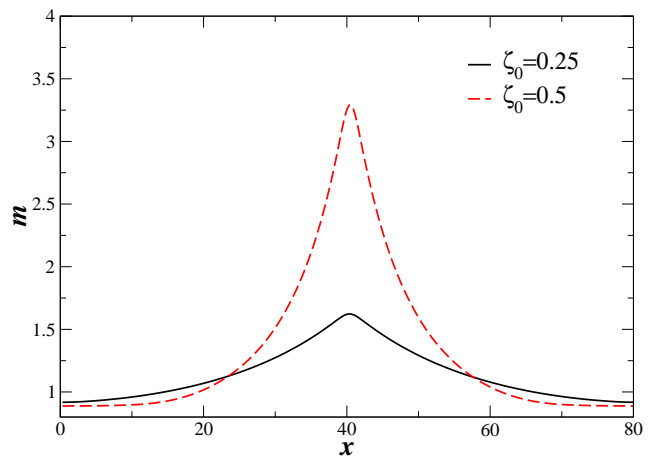}
\caption{Motor contrectation profiles  $m$ for different values of
$\zeta_0$ for isolated aster for $H=0.125,B=0.06, D=5$, and
$m_0=1, \rho_0=4$. } \label{fig:motor1}
\end{figure}

\begin{figure}[ptb]
\includegraphics[width=1.68in,angle=0]{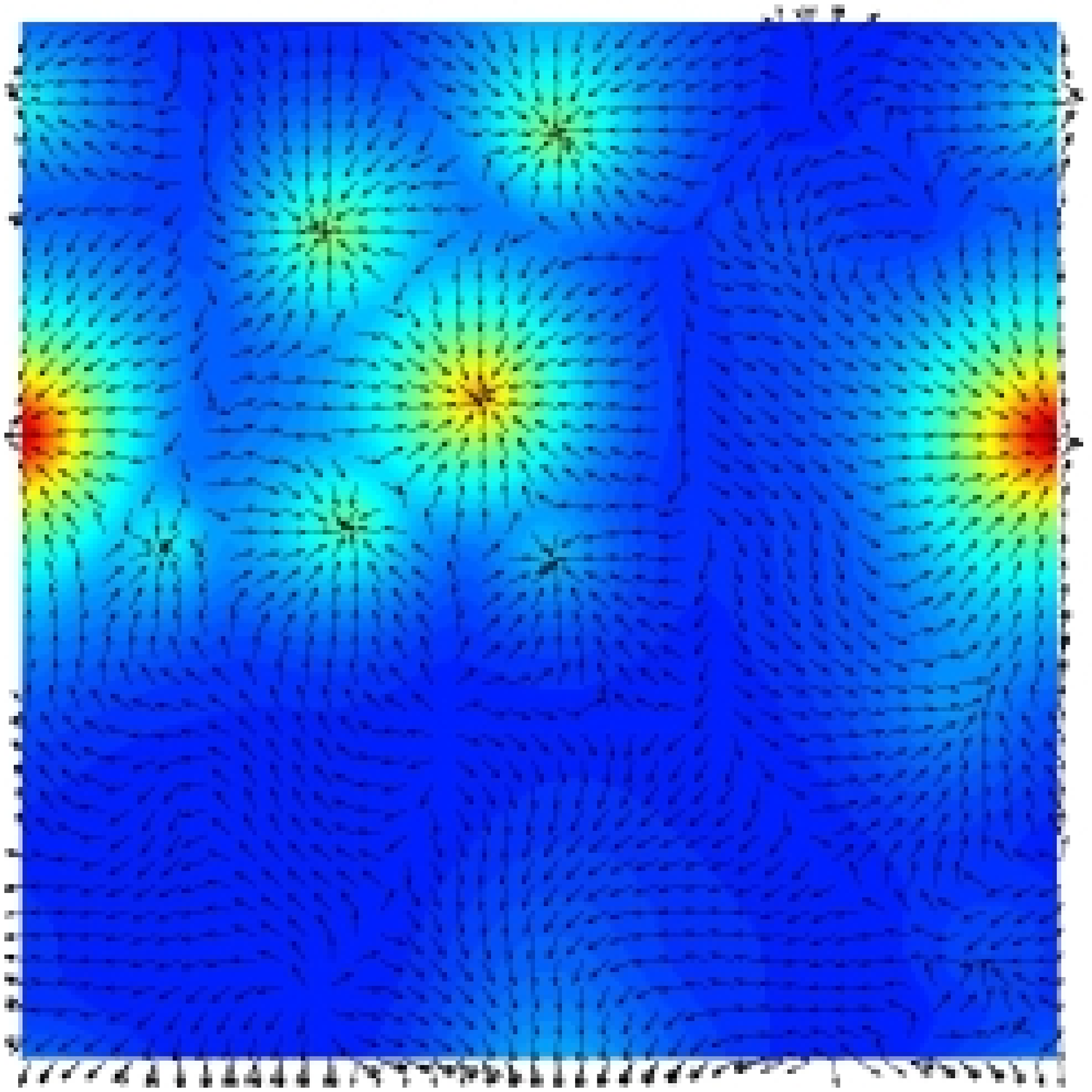}
\includegraphics[width=1.68in,angle=0]{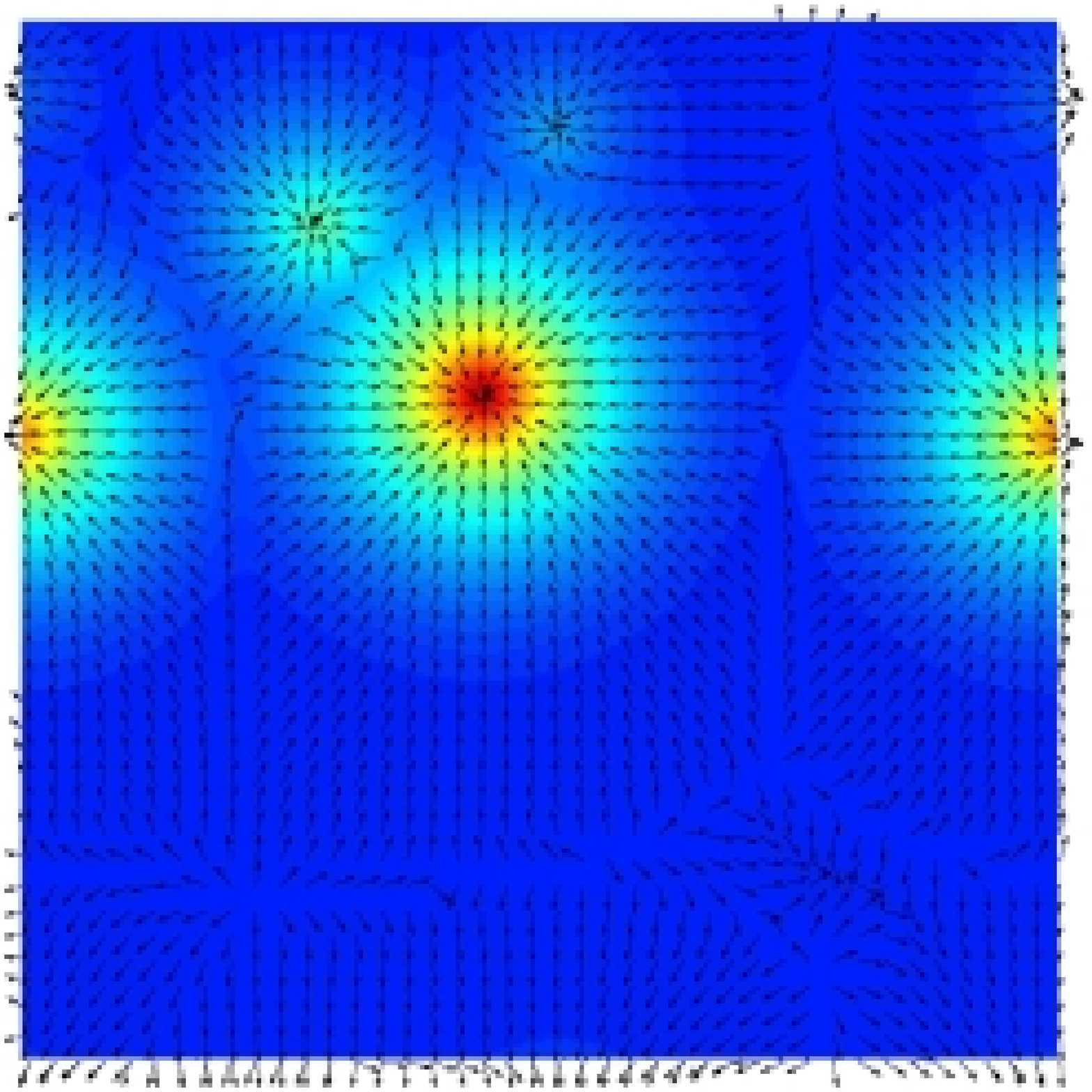}
\caption{Color-coded images of density of motors at $t=200$ (left)
and $t=1000$ (right), red corresponds to maximum density and blue
to minimum density, domain size is $80\times80$ units,
$\zeta_0=0.5$, $H=0.125$, $D=5$, $m_0=1$. Arrows show orientation
of microtubules.} \label{fig:mot}
\end{figure}

Selected results are shown in Fig. \ref{fig:motor1},\ref{fig:mot}.
In agreement with experiment, we observed that motors tend to
accumulate in the center of an aster or a vortex, see
Fig.\ref{fig:motor1},\ref{fig:mot}. Otherwise, qualitative
behavior of formation of asters and vortices remains the same. As
seen in Figure \ref{fig:mot}, initial multi-aster
state coarsens and leads to the formation of a network of large
asters separated by the domain walls.

\section{Conclusions}

In this paper we derived continuous equations for the evolution of
microtubule concentration and orientation due to their interaction
via molecular motors.  We found that an initially disordered
system exhibits an ordering instability qualitatively similar to
the nematic phase transition in ordinary polymers at high density.
The important difference is that here the ordering instability is
mediated by molecular motors and can occur at arbitrary low
densities of microtubules. At the nonlinear stage, the instability
leads to the experimentally observed formation of asters and
vortices. Similar vortices were observed in a system of
interacting granular rods \cite{kudrolli,at03}. While we find that
it suffices to consider only the density of tubules to explain the
basic phenomenology, a better agreement with experiment is
obtained when we include variable motor density and the motors
attachment to the substrate \cite{nedelec01,lee01,kim03}.

Many aspects of self-assembly in tubule-motor systems require further inverstigation. In
particular, we anticipate that flexibility of the microtubules may
have a strong effect on the details of interaction, the question
which we plan to address in future work. Another interesting
question is role of the hydrodynamic interaction between the
microtubules and effects of fluctuations on the orientation
transition. Furthermore, our theory is derived in the limit of small
density of the microtubules and takes into account only binary
interactions among tubules.
Certainly, multiparticle interactions are important in the high
density state, such as bundles. Generalization of our work to the
multiparticle interaction is a very challenging problem.

There are many  predictions following from our analysis which possibly
deserve experimental verification. For example, we find that the
anisotropy of the interaction kernel is associated with the
inhomogeneous density of motors along the microtubules. We observed that
the motors attached to the substrate reduce the density of microtubules
in the cores of asters and vortices. We predicted an acceleration
instability which leads to a drift of isolated asters. Hopefully, new
generations of experiments will be able to address these issues.

We thank Jacques Prost, Anthony Maggs,  Leo Kadanoff and Valerii
Vinokur for useful discussions. This work was supported by the
U.S. Departemtn of Energy, grants W-31-109-ENG-38 (IA) and DE-FG02-04ER46135 (LT).

\appendix

\section{Continuum equations for arbitrary $\gamma, \phi_0$}
\label{app1}
\begin{widetext}
We assume that two rods interact when angle between them is less
or equal $\phi_0$ independent of their
spatial location. The master equation reads
\begin{eqnarray}
&&\frac{\partial P(\phi,t)}{\partial t}=
D_r\frac{\partial^2P(\phi,t)}{\partial \phi^2}
\nonumber\\
&& +g \int_0^{\phi_0/2}\int_{0}^{2\pi} 2
dydxP(x+y,t)P(x-y,t)\left[\delta(\phi-\gamma(x-y)-(1-\gamma)(x+y))-\delta(\phi-x-y)\right]
\nonumber\\
&&+g \int_{\pi-\phi_0/2}^\pi\int_{0}^{2\pi} 2
dydxP(x+y,t)P(x-y,t)\left[\delta(\phi-\gamma(x-y)-(1-\gamma)(x+y+2\pi))-\delta(\phi-x-y)\right]
\label{master_a1}
\end{eqnarray}
where $x=(\phi_1+\phi_2)/2, y=(\phi_2-\phi_1)/2$ (take into
account that $d \phi_1 d \phi_2 = 2 dx dy $), and $g$ is the
collision rate,  and we also included thermal diffusion of rods
orientation $\propto D_r$.

Performing integration over $x$ from 0 to $2\pi$, we get
\begin{eqnarray}
\frac{\partial P(\phi,t)}{\partial t}=
D_r\frac{\partial^2P(\phi,t)}{\partial \phi^2}+ 2 g
\int_0^{\phi_0/2}dy\left[P(\phi+2\gamma y,t)
P(\phi+2(\gamma-1)y,t)-P(\phi,t)P(\phi-2y,t)\right]
\nonumber\\
+2 g \int_{\pi-\phi_0/2}^\pi dy\left[P(\phi+2\gamma
y+2\pi\gamma,t)P(\phi+2(\gamma-1)y+2\pi\gamma,t) -
P(\phi,t)P(\phi-2y,t)\right] \label{master2_a1}
\end{eqnarray}

Changing variable $y \to y+\pi$ in the second term in
(\ref{master2_a1}), we obtain
\begin{eqnarray}
\frac{\partial P(\phi,t)}{\partial t}=
D_r\frac{\partial^2P(\phi,t)}{\partial \phi^2}+2 g
\int_{-\phi_0/2}^{\phi_0/2}dy\left[P(\phi+2y\gamma,t)P(\phi+2y(\gamma-1),t)
-P(\phi,t)P(\phi-2y,t)\right] \label{master2a_a1}
\end{eqnarray}

The case when all rods interact corresponds to $\phi_0=\pi$, and
Eq.(\ref{master2a_a1}) simplifies to
\begin{eqnarray}
\frac{\partial P(\phi,t)}{\partial t}+g P(\phi,t)=
D_r\frac{\partial^2P(\phi,t)}{\partial \phi^2}+ 2 g
\int_{-\pi/2}^{\pi/2}dy[P(\phi+2y\gamma,t)P(\phi+2y(\gamma-1),t)
\label{master2b_a1}
\end{eqnarray}

By substitution $y \to w/2$ Eq. (\ref{master2a_a1}) can be
transformed into the form
\begin{eqnarray}
\frac{\partial P(\phi,t)}{\partial
t}-\frac{\partial^2P(\phi,t)}{\partial \phi^2}=
\int_{-\phi_0}^{\phi_0}dw\left[P(\phi+w\gamma,t)P(\phi+w(\gamma-1),t)
-P(\phi,t)P(\phi-w,t)\right] \label{master2c_a1}
\end{eqnarray}
where we changed variables $t \to D_r t$, $P \to  g P/D_r $.

Let us consider a Fourier expansion of the probability distribution
\begin{equation}
P(\phi,t)=\sum_{k=-\infty}^{\infty} P_k(t) e^{ik\phi}
\label{Pk_a1}
\end{equation}
where $P_{-k}=P_k^*$. The Fourier harmonics $P_k$ are given by
angular averages of $\exp(ik\phi)$, see Eq. (\ref{cumm}). The constant
zeroth harmonic $P_0=1/2\pi \rho$, where $\rho$ is the number
density,
\begin{equation}
\rho=\int _0^{2 \pi} d \phi  P(\phi, {\bf r})= 2 \pi P_0
\label{dens_a1}
\end{equation}
and the real and imaginary parts of $P_1$ represent the components
of the orientation vector $ {\bf \tau} = (\langle \cos \phi\rangle,
\langle \sin \phi\rangle)$.  Accordingly, $\tau_x+i
\tau_y= \langle \exp [ i \phi] \rangle = P_{1}^*$.

After substitution of (\ref{Pk_a1}) into Eq.(\ref{master2c_a1})
we obtain the infinite series of equations for $P_k$
\begin{equation}
\dot P_k + k^2P_k=2\phi_0\sum_n\sum_m P_n P_m
(S[\phi_0(n\gamma+m(\gamma-1))] - S(m\phi_0)]\delta_{n+m,k}
\label{Pk1_a1}
\end{equation}
where $S(x)=\sin x/x$, and $\delta_{n+m,k}$ is the Kroneker symbol.

For $\phi_0=\pi$, the latter equation simplifies to
\begin{equation}
\dot P_k +(k^2+1)P_k=2\pi \sum_n\sum_m P_n P_m
S[\pi(n\gamma+m(\gamma-1)]\delta_{n+m,k} \label{Pk1a_a1}
\end{equation}

Now we have to truncate this series. Assuming $P_n=0$  for all
$|n|>2$, from Eq.(\ref{Pk1_a1}) one obtains $\dot P_0= 0$,
\begin{eqnarray}
&&\dot P_1 + P_1= P_0 P_1 2\phi_0 \left[S[\phi_0(\gamma-1)]+
S[\phi_0\gamma]-S(\phi_0)-1\right] \nonumber \\
&+&2 \phi_0 P_2 P_1^* \left[S[\phi_0(\gamma+1)]+
S[\phi_0(\gamma-2)]-S(2\phi_0)-S(\phi_0)\right] \label{Pk2_a1}
\end{eqnarray}
and
\begin{eqnarray}
\dot P_2 +4 P_2= P_0 P_2 2\phi_0 \left[S[2\phi_0(\gamma-1)]+
S[2\phi_0\gamma]-S(2\phi_0)-1\right] + 2 \phi_0
P_1^2\left[S[\phi_0(2\gamma-1)] -S(\phi_0)\right] \label{Pk3_a1}
\end{eqnarray}

Neglecting the time derivative $\dot P_2$, we obtain $P_2=AP_1^2$
with
\begin{equation}
A=\frac{S[\phi_0(2\gamma-1)] -S(\phi_0)}{ 2 /  \phi_0 -
(S[2\phi_0(\gamma-1)]+S[2\phi_0\gamma]-S(2\phi_0)-1)\rho/2\pi}
\label{A}
\end{equation}
That allows us to close the equation for $P_1$,
\begin{eqnarray}
&&\dot P_1 +P_1=  \rho P_1
\phi_0\pi^{-1}\left[S[\phi_0(\gamma-1)]+
S[\phi_0\gamma]-S(\phi_0)-1\right]
\nonumber\\
&&+ 2A \phi_0 |P_1|^2P_1  \left[S[\phi_0(\gamma+1)]+
S[\phi_0(\gamma-2)]-S(2\phi_0)-S(\phi_0)\right] \label{Pk2a_a1}
\end{eqnarray}

For $\phi_0=\pi$,
\begin{eqnarray}
\dot P_1 +(1 + \rho)P_1= \rho P_1\left[S[\pi(\gamma-1)]+
S[\pi\gamma]\right] +2 \pi A_0 |P_1|^2P_1 \left[S[\pi(\gamma+1)]+
S[\pi(\gamma-2)]\right] \label{Pk2b_a1}
\end{eqnarray}
where
\begin{equation}
A_0=\frac{2\pi S[\pi(2\gamma-1)]} {4   -(S[2\pi(\gamma-1)]+
S[2\pi\gamma]-1)\rho}
\end{equation}
As seen from this equation, for $0<\gamma<1$ we obtain an ordering
instability which for large enough $\rho>\rho_{c}$ leads to a
spontaneous alignment of filaments.

\section{Evaluation of terms $Z_0,Z_1$}
\label{app2}
\subsection{Isotropic term $Z_0$.}
We introduce new variable $\bxi=\bf r-\bf r_1$, and obtain after simple algebra
\begin{eqnarray}
&& Z_0= \int d\bxi \int_{-\phi_0}^{\phi_0}dw W_0(|\bxi|)
\left[P({\bf r}-\bxi/2 ,\phi+w\gamma,t)P({\bf r}+\bxi/2,
\phi+w(\gamma-1),t) \right. \left. -P({\bf r},\phi,t)P({\bf
r}-\bxi ,\phi-w,t)\right] \label{master5b}
\end{eqnarray}

Now we assume that the probability distributions are smooth
functions of spatial coordinates on the scale of the rod length
$l$, and expand them near ${\bf r}$,
\begin{equation}
P({\bf r}+\bxi,\phi,t)=P({\bf r},\phi,t)+(\bxi\cdot \nabla)P({\bf
r},\phi,t)+{1\over 2}(\bxi\cdot \nabla)^2P({\bf
r},\phi,t)+O(\xi^3), \label{expansion}
\end{equation}
Performing integration over $\bxi$ using kernel (\ref{W5}), we get
\begin{eqnarray}
&& Z_0 = \int_{-\phi_0}^{\phi_0}dw \left[P({\bf
r},\phi+w\gamma,t)P({\bf r}, \phi+w(\gamma-1),t) -P({\bf
r},\phi,t)P({\bf r},\phi-w,t)\right]
\nonumber\\
&& + {b^2\over 16} \int_{-\phi_0}^{\phi_0}dw \left[-2\nabla P({\bf
r} ,\phi+w\gamma,t)\nabla P({\bf r}, \phi+w(\gamma-1),t)\right.
\\
&&+\nabla^2 P({\bf r},\phi+w\gamma,t) P({\bf r},
\phi+w(\gamma-1),t) +P({\bf r},\phi+w\gamma,t)\nabla^2P({\bf r},
\phi+w(\gamma-1),t) \left.-4P({\bf r},\phi,t)\nabla^2 P({\bf
r},\phi-w,t)\right] \nonumber
\label{master5c}
\end{eqnarray}
Now we expand $P$'s in the Fourier series over $\phi$, $P({\bf
r},\phi,t)=\sum_n P_n e^{in\phi}$. The $k$th Fourier component of
$Z_0$ reads
\begin{eqnarray}
&&Z_0^k= 2\phi_0\sum_n\sum_m\delta_{n+m,k} P_n P_m
[S[\phi_0(n\gamma+m(\gamma-1))] - S(m\phi_0)]
\nonumber\\
&& +{b^2\phi_0\over 8}\sum_n\sum_m \delta_{n+m,k}[(-2\nabla P_n
\nabla P_m +\nabla^2 P_n P_m +P_n \nabla^2
P_m)S[n\gamma\phi_0+m(\gamma-1)\phi_0]
 - 4P_n\nabla^2 P_m S(m\phi_0)] \label{master5d}
\end{eqnarray}
The first sum in Eq.(\ref{master5d}) coincides with the spatially
uniform case (\ref{Pk1_a1}). For $k=0$ we obtain (keeping only
terms up to $|n|,|m|=1$)
\begin{eqnarray}
&&Z_0^0= -{b^2\phi_0\over 8} [\nabla^2(P_0^2)
+2\nabla^2(|P_1|^2)S(\phi_0)] \label{master5e}
\end{eqnarray}
For $k=1$
\begin{eqnarray}
&&Z_0^1= 2 \phi_0P_0P_1\left[S[\phi_0(\gamma-1)]+
S[\phi_0\gamma]-S(\phi_0)-1\right]
+2\phi_0P_2P_1^*\left[S[\phi_0(\gamma+1)]+
S[\phi_0(\gamma-2)]-S(2\phi_0)-S(\phi_0)\right]
\nonumber\\
&& +{b^2\phi_0\over 8} [(\nabla^2P_0P_1 +P_0\nabla^2P_1-2\nabla
P_0\nabla P_1)(S[\gamma\phi_0]+S[(\gamma-1)\phi_0])
-4P_0\nabla^2P_1S[\phi_0]-4P_1\nabla^2P_0]
\nonumber\\
&&+{b^2\phi_0\over 8} [(-2\nabla P_1^* \nabla P_2 +\nabla^2 P_1^*
P_2 +P_1^*\nabla^2P_2)S[(2-\gamma)\phi_0]+ \ (-2\nabla P_2 \nabla
P_1^* +\nabla^2 P_2 P_1^* +P_2\nabla^2P_1^*)S[(1+\gamma)\phi_0]
\nonumber\\
&& -4P_1^*\nabla^2P_2S(2\phi_0)-4P_2\nabla^2P_1^*S(\phi_0)]
\label{master5f}
\end{eqnarray}
[this equations derived with the aid of Mathematica \cite{mathematica}].

We can again use the expression $P_2=AP_1^2$ with constant $A$
obtained for the spatially uniform case (\ref{A}).

Now, if we set $\phi_0=\pi$, $\gamma=1/2$ and neglect higher order
terms in the differential operators, we obtain
\begin{eqnarray}
Z_0^0 &=& -{b^2\pi\over 8}  \nabla^2(P_0^2)  \nonumber  \\
Z_0^1 &=& 2 P_0 P_1(4-\pi)-{8\over 3} P_2P_1^*
 +{b^2 \over 2} \left [ \nabla^2 (P_0P_1 ) -4\nabla
P_0\nabla P_1-\pi P_1\nabla^2 P_0 - ( \nabla^2 (P_2 P_1^*
)-4\nabla P_1^* \nabla P_2)/3
  \right ]\nonumber
 \label{interm1}
\end{eqnarray}

\subsection{Anisotropic term $Z_1$.}
For compactness of
notations we introduce the following definitions $z=x+i y,  \psi =
n_x+i n_y = \exp [i \phi]$. Scalar product assumes the form $ {\bf
r} \cdot {\bf n}= Re( z^* \exp[i \phi])$. Now Eq. (\ref{master4c})
can be written in the form
\begin{eqnarray}
&&Z_1 =  \int d{\bf r_1} \int_{-\phi_0}^{\phi_0}dw \nonumber
\\
&& \left[ 2 W_0(2 ({\bf r}_1-{\bf r}))Re(2  ( z_1-
z)^*(e^{i(\phi+\gamma w)}-e^{i (\phi+(\gamma-1)w)}) P({\bf
r}_1,\phi+\gamma w,t)P(2 {\bf r}-{\bf
r}_1,\phi+(\gamma-1)w,t)\right.
\nonumber\\
&& \left.- W_0({\bf r}_1-{\bf r})Re((z_1- z)^* (e^{i(\phi-w)}-e^{i
\phi})) P({\bf r},\phi,t)P({\bf r}_1,\phi-w,t)\right]
\label{master6b}
\end{eqnarray}
Let us introduce $\bxi=2({\bf r_1}-{\bf r})$, $\zeta=2(z_1- z)$ in
the first integrand of Eq.(\ref{master6b}) and $\bxi={\bf
r_1}-{\bf r}$, $\zeta=z_1- z$ in the second, and obtain
\begin{eqnarray}
&Z_1 &=  \int d\bxi \int_{-\phi_0}^{\phi_0}dw W_0( |\bxi|)
\nonumber
\\
&& \left[  Re( \zeta^*(e^{i(\phi+\gamma w)}-e^{i
(\phi+(\gamma-1)w)}) P({\bf r}+\bxi/2,\phi+\gamma w,t)P({\bf
r}-\bxi/2,\phi+(\gamma-1)w,t)\right.
\nonumber\\
&& -\left. Re(\zeta^* (e^{i(\phi-w)}-e^{i \phi})) P({\bf
r},\phi,t)P({\bf r}+\bxi,\phi-w,t)\right] \label{master7b}
\end{eqnarray}
Now we again make use of the Fourier expansion $P({\bf
r},\phi,t)=\sum_n P_n({\bf r},t)e^{in\phi}$. Multiplying
Eq.(\ref{master7b}) by $(2\pi)^{-1}e^{-ik\phi}$ and integrating
over $\phi$ from 0 to $2\pi$ , we get
\begin{eqnarray}
&&Z_1^k = {1\over 2\pi} \int d\bxi\int_0^{2\pi}d\phi
\int_{-\phi_0}^{\phi_0}dw W_0(|\bxi|)e^{-ik\phi} \nonumber
\\
&& \left[Re(\zeta^*(e^{i(\phi+\gamma w)}-e^{i (\phi+(\gamma-1)w)})
P({\bf r}+\bxi/2,\phi+\gamma w,t)P({\bf
r}-\bxi/2,\phi+(\gamma-1)w,t)\right.
\nonumber\\
&& -\left. Re(\zeta^* (e^{i(\phi-w)}-e^{i \phi})) P({\bf
r},\phi,t)P({\bf r}+\bxi,\phi-w,t)\right] \label{master7c}
\end{eqnarray}
Using periodicity in $\phi$ we shift variables $\phi\to\phi-\gamma
w$ in the first integrand:
\begin{eqnarray}
&&Z_1^k = {1\over 2\pi} \int d\bxi\int_0^{2\pi}d\phi
\int_{-\phi_0}^{\phi_0}dw W_0(|\bxi| )e^{-ik\phi}
Re(\zeta^*(e^{i\phi}-e^{i (\phi-w)})
\nonumber \\
&& \left[P({\bf r}+\bxi/2,\phi,t)P({\bf
r}-\bxi/2,\phi-w,t)e^{ik\gamma w} +P({\bf r},\phi,t)P({\bf
r}+\bxi,\phi-w,t)\right] \label{master7d}
\end{eqnarray}
Changing variables $\phi\to\phi+w$ and then $w\to -w$ where
appropriate, we obtain
\begin{eqnarray}
&&Z_1^k = {1\over 2\pi} \int d\bxi\int_0^{2\pi}d\phi
\int_{-\phi_0}^{\phi_0}dw W_0(|\bxi|)e^{-ik\phi}
Re(\zeta^*e^{i\phi}) \nonumber
\\
&& \left[P({\bf r}+\bxi/2,\phi,t)P({\bf
r}-\bxi/2,\phi-w,t)e^{ik\gamma w} -P({\bf
r}+\bxi/2,\phi-w,t)P({\bf
r}-\bxi/2,\phi,t)]e^{ik(1-\gamma)w}\right.
\nonumber\\
&& \left.+P({\bf r},\phi,t)P({\bf r}+\bxi,\phi-w,t)- P({\bf
r},\phi-w,t)P({\bf r}+\bxi,\phi,t)e^{ikw}\right] \label{master7a}
\end{eqnarray}
Let us consider again the zeroth and first moments only. For $Z_1^0$ we
get
\begin{eqnarray}
&&Z_1^0 = {1\over 2\pi} \int d\bxi\int_0^{2\pi}d\phi
\int_{-\phi_0}^{\phi_0}dw W_0(|\bxi|) Re(\zeta^*e^{i\phi})
\nonumber
\\
&& \left[P({\bf r}+\bxi/2,\phi,t)P({\bf r}-\bxi/2,\phi-w,t)
-P({\bf r}+\bxi/2,\phi-w,t)P({\bf r}-\bxi/2,\phi,t)]\right.
\nonumber\\
&& \left.+P({\bf r},\phi,t)P({\bf r}+\bxi,\phi-w,t)- P({\bf
r},\phi-w,t)P({\bf r}+\bxi,\phi,t)\right] \label{master7-c}
\end{eqnarray}
After cumbersome  transformations in Mathematica it can
be reduced to
\begin{eqnarray}
&&Z_1^0 = { 1\over 64\pi  } \int_0^{2\pi}d\phi
\int_{-\phi_0}^{\phi_0}dw  \left[ 3 Re \left ( \bar \nabla^*
\exp(i\phi) \right) \left( P(\phi) \Delta P(\phi-w) - P(\phi-w)
\Delta P(\phi) \right)
\right. \nonumber \\
&&- \left. 2 Im  \left ( \bar \nabla^* \exp(i\phi) \right) \left(
P(\phi)_y P(\phi-w)_x - P(\phi)_x P(\phi-w)_y \right)
  \right] \label{master7-d}
\end{eqnarray}
(here $\bar \nabla=\partial_x+i \partial_y, \bar
\nabla^*=\partial_x-i \partial_y$). Performing integration over
$\phi$ and $w$, we obtain
\begin{eqnarray}
&&Z_1^0 = { \pi\over  32 } \left[ 3   \nabla^* \left( P_{-1}
\Delta P_0 - P_0 \Delta P_{-1} \right)-
 {2 \over i}    \left( \bar \nabla^*  \left(
\partial_y P_{-1}   \partial_x P_0 - \partial_x P_{-1} \partial_y P_0 \right) \right)
  \right] +c.c.\label{master7-e}
\end{eqnarray}
 This expression can be written in the vector form
\begin{eqnarray*}
&&Z_1^0={ \pi\over  16 } \left[ 3   \nabla \cdot
\left( {\bf \tau}  \Delta \rho - \rho  \Delta {\bf \tau} \right)
-2 \frac{\partial}{\partial x} \left(
\partial_y \tau_y  \partial_x \rho - \partial_x \tau_y  \partial_y \rho \right)
+ 2 \frac{\partial}{\partial y} \left(
\partial_y \tau_x  \partial_x \rho - \partial_x \tau_x  \partial_y \rho \right)
  \right] \label{master7-g}
  \end{eqnarray*}

Similarly, after some algebra we also obtain the anisotropic part of the
first moment, which for $\gamma=1/2$ reads
\begin{eqnarray}
&&Z_1^1 = { b^2\over 4\pi} \int_0^{2\pi}d\phi
\int_{-\phi_0}^{\phi_0}dw e^{-i\phi}
\left[[Re(e^{i\phi}\bar\nabla^*) P({\bf r},\phi,t)P({\bf
r},\phi-w,t)-P({\bf r},\phi,t)Re(e^{i\phi}\bar\nabla^*) P({\bf
r},\phi-w,t)]e^{i w/2} \right. \nonumber
\\
&& \left.+P({\bf r},\phi,t)Re(e^{i\phi}\bar\nabla^*) P({\bf
r},\phi-w,t)- P({\bf r},\phi-w,t)Re(e^{i\phi}\bar\nabla^*) P({\bf
r},\phi,t)e^{iw}\right] \label{master7_1i}
\end{eqnarray}
Eq. (\ref{master7_1i}) can be written as
\begin{eqnarray}
&&Z_1^1 = { b^2\over 4\pi} \int_0^{2\pi}d\phi
\int_{-\phi_0}^{\phi_0}dw e^{-i\phi} (1- e^{i w/2})  \nonumber
\\
&& \left[  e^{i w/2}  Re(e^{i\phi}\bar\nabla^*) P({\bf
r},\phi,t)P({\bf r},\phi-w,t)+ P({\bf
r},\phi,t)Re(e^{i\phi}\bar\nabla^*) P({\bf r},\phi-w,t)  \right]
\label{master7_2i}
\end{eqnarray}
Now performing integration over $\phi,w$ we obtain
\begin{eqnarray}
&&Z_1^1 = {\phi_0  b^2\over 2   } \sum_{m=0}^{\infty}\left[ \left(
S((1/2-m)\phi_0)-S((1-m) \phi_0) \right) \left( \bar\nabla^*
P_{-m}
P_m + \bar\nabla P_{2 - m} P_m \right) \right. \nonumber \\
&& \left. + \left( S(m\phi_0)-S((1/2-m) \phi_0) \right) \left(
\bar\nabla^* P_{m} P_{-m} + \bar\nabla P_{m} P_{2-m} \right)
\right] \label{master7_3i}
\end{eqnarray}
Substituting $\phi_0=\pi$, and keeping only the first two moments, we
obtain
\begin{eqnarray}
Z_1^1 = {\pi  b^2\over 2  } \left[ \bar\nabla^* P_{0}
P_{0}-P_1\left( \bar\nabla^* P_{-1} + \bar\nabla P_{1}
\right)+S_{1/2} \left(\bar\nabla^*P_{-1}P_1-\bar\nabla^* P_{1}
P_{-1}\right)   + S_{3/2} \left( \bar\nabla^* P_{1} P_{-1}
-\bar\nabla^* P_{-1} P_{1} \right) \right] \label{master7_4i}
\end{eqnarray}
where $S_{1/2}=S(\pi/2)$ and $S_{3/2}=S(3\pi/2)$.

After some transformations we obtain
\begin{eqnarray}
Z_1^1 = {\pi b^2\over 2  } \left[ \bar\nabla^* P_{0} P_{0}-
(1-S_{1/2}+S_{3/2})\left( \bar\nabla^* P_{-1} P_{1} + \bar\nabla
P_{1} P_{1} \right)  -(S_{1/2}-S_{3/2}) \left( \bar\nabla P_{1}
P_{1}+\bar\nabla^* P_{1} P_{-1} \right) \right] \label{master7_6i}
\end{eqnarray}
In the vector form it gives
\begin{eqnarray}
&& Z_1^1={\pi  b^2\over 2  } \left[ {1\over
4\pi^2}\rho\nabla \rho - 2(1-S_{1/2}+S_{3/2}) {\bf \tau } \nabla
\cdot \tau
 -2 (S_{1/2}-S_{3/2}) ({\bf \tau}  \nabla) {\bf \tau}   \right]     \label{master7_7i}
\end{eqnarray}

Since $S_{1/2}=2/\pi, S_{3/2}=-2/3 \pi$, we obtain
\begin{eqnarray}
&&Z_1^1 =  b^2 \left[\frac{1}{8 \pi} \rho\nabla\rho-
\left(\pi-{8\over 3}\right) \btau(\nabla\cdot\btau)- {8\over 3}
({\bf \tau}  \nabla) \btau \right] \label{master7_5i}
\end{eqnarray}

\end{widetext}


\begin{thebibliography}{10}
\bibitem{friction} In the three-dimensional case,  the drag
coefficients  $\xi_\parallel$, $\xi_\perp$ for rod-like objects
are different by the factor of two \cite{doi}. However, the
relative  difference between $\xi_\parallel$ and  $\xi_\perp$
decreases in thin fluid films and membranes \cite{levine}.
\end{thebibliography}
\end{document}